# Materials design criteria for ultra-high thermoelectric power factors in metals

Patrizio Graziosi[1], Kim-Isabelle Mehnert,[2] Rajeev Dutt[3,] Jan-Willem G. Bos[2], and Neophytos Neophytou[3*]

[1] CNR – ISMN, via Gobetti 101, 40129, Bologna, Italy

[2] EaStCHEM School of Chemistry, University of St Andrews, North Haugh, St Andrews, KY16 9ST, UK.

[3] School of Engineering, University of Warwick, Coventry, CV4 7AL, UK

[*] n.neophytou@warwick.ac.uk

## Abstract

Metals have high electronic conductivities, but very low Seebeck coefficients, which traditionally make them unsuitable for thermoelectric materials. Recent studies, however, showed that metals can deliver ultra-high thermoelectric power factors (PFs) under certain conditions. In this work, we theoretically examine the electronic structure and electronic transport specifications which allow for such high PFs. Using Boltzmann transport (BTE) simulations and a multi-band electronic structure model, we show that metals with: i) high degree of transport asymmetry between their bands, ii) strong inter-band scattering, and iii) a large degree of band overlap, can provide ultra-high power factors. We show that each of these characteristics adds to the steepness of the transport distribution function of the BTE, which allows for an increase of the Seebeck coefficient to sizable values, *simultaneously* with an increase in the electrical conductivity. This work generalizes the concept that transport asymmetry (i.e., mixture of energy regions of high and low contributions to the electrical conductivity), through a combination of different band masses, scattering strengths, or energy filtering scenarios, etc., can indeed result in very high thermoelectric power factors, even in the absence of a material bandgap. Under certain conditions, transport asymmetry can over-compensate any performance degradation to the PF due to bipolar conduction and the naturally low Seebeck coefficients that otherwise exist in this class of materials.

I. INTRODUCTION

Thermoelectric generators (TEGs), which convert the heat flow arising from temperature gradients directly into electricity and vice versa, can play a significant role in the search for sustainable paths for energy harvesting and cooling in a variety of applications. However, large scale exploitation has been limited by the high prices, toxicity, scarcity, and low efficiencies of the prominent thermoelectric (TE) materials [1,2]. The TE performance is quantified by the figure of merit $ZT = \sigma S^2 T/(\kappa_e + \kappa_L)$, where $\sigma$ is the electrical conductivity, $S$ is the Seebeck coefficient, $T$ is the absolute temperature, and $\kappa_e$ and $\kappa_L$ are the electronic and lattice parts of the thermal conductivity, respectively. The product $\sigma S^2$ is called the power factor (PF). Over the last two decades, progress on TE materials has been rapidly expanding with the synthesis of a myriad of new materials and their alloys. While the thermal conductivity reduction is mainly addressed by nanostructuring which increases phonon scattering, the PF is determined by the complex features that new materials exhibit in their electronic structure [1–4]. Traditionally, the optimization of the PF in TE semiconductors focuses on doping and the study of the role of valley pockets and band alignment [5–14], highlighting the need for low inter-valley scattering strength between the different valleys to allow high conductivity [15–18].

Thermoelectric materials are typically semiconductors with a sizable bandgap. The Seebeck coefficient depends on the transport asymmetry around the Fermi level, $E_F$, (i.e. the variation in the contribution to transport from electrons with energies above/below $E_F$). Thus, most metals (typically close to zero asymmetry) provide very small Seebeck coefficients and PF, and are typically avoided. Low bandgap materials also suffer from bipolar transport, which increases the electronic thermal conductivity, but also reduces the Seebeck coefficient since the conduction and valence bands have Seebeck coefficients of opposite sign [19,20]. Nevertheless, some low gap materials and semimetals are still considered for thermoelectric applications [21–23]. Recently, we have also proposed that low-doped narrow gap semiconductors with asymmetric bandstructures and asymmetric transport features between the conduction and valence bands, can deliver very high PFs [14,24]. In recent years, new classes of materials such as topological semimetals also indicate promising TE performance [25]. But in general TE materials based on metals and low bandgap materials are considered as not promising due to their very low Seebeck coefficients.

On the other hand, utilizing highly conductive metallic, semi-metallic, or highly doped materials that have very high conductivity, while engineering a transport asymmetry around the Fermi level to obtain finite Seebeck values, has been discussed in different settings in the past. For example, Karamitaheri *et al.* suggested that extended defects and surface roughness in metallic graphene nanoribbons can create a transport asymmetry and a so-called 'transport gap', leading to a finite Seebeck coefficient and substantial improvement in TE performance [26]. Energy filtering is another way to introduce transport asymmetry between low and high energy carriers. This is typically engineered by the introduction of energy barriers to filter low energy carriers [27-30]. It was also shown theoretically and experimentally that it is much more efficient when applied to ultra-highly doped materials (close to metallic), for which the Seebeck coefficient is very low to begin with [31]. Certain impurities can also create resonant states in some materials which create asymmetry in the bandstructure at high energies, and improve the Seebeck coefficient and the PF [7, 32]. Recently, Garmroudi *et al* have experimentally demonstrated that highly conducting 'good' metals such as $Ni_xAu_{1-x}$ alloys [33, 34], can have very high PFs due to the large density of states asymmetry at the transition between the *s*-orbitals of Au and the *d*-orbitals of Ni. The PF benefits of the *s-d* orbital scattering in metals were known since the 1930s, but were overlooked for decades in favor of semiconductors. All of the aforementioned works, rather unconventionally, attempt to induce a finite (albeit small) Seebeck in highly conductive materials, instead of increasing the conductivity of materials with high Seebeck coefficient to begin with. Note, however, that these are not the first time metals are reported to exhibit large PFs. Intermediate valence or heavy fermion intermetallic compounds such as $YbAl_3$ or $CePd_3$ (which also exhibit the Kondo effect at low temperatures), have demonstrated some of the highest PFs around room temperature (in some cases >15 mW/mK$^2$) [35, 36, 37, 38]. This is around three times the room temperature PF of $Bi_2Te_3$, arising from a substantial value of the Seebeck coefficient (compared to other metals), together with their high metallic conductivity.

Motivated by these demonstrations, and especially the work of Garmroudi *et al*. [33], here we perform a theoretical investigation using the Boltzmann Transport Equation (BTE) to unveil the necessary conditions upon which such high PFs can be achieved in metals. We consider a simple model of overlapping parabolic bands to mimic metallic behavior. We show that: i) the degree of band transport asymmetry (i.e. the differences between the contribution of each band to electronic transport), ii) the degree of inter-band scattering, and iii) the degree of band overlap, are three parameters that determine this behaviour. The overlap and inter-band scattering between asymmetric bands leads to an effective filtering (by scattering) of charge carriers in the overlap region and creates crucial sharp features in the energy dependence of the transport distribution function of the BTE [39–42]. Such behavior can be observed in cases of alloys between transition metals [33,43,44], materials with a combination of flat and dispersive bands [4,7,15,33,45,46], and could also be observed in materials with an overlap between the CB and VB, having the so-called *negative bandgap*, which could be observed in the Heusler alloy material group among others [47–49]. Note that studies on metals and semimetal TEs can be found sporadically in the literature [50], but this work presents a systematic investigation of the TE properties of asymmetric bands in metals and investigates the energy filtering caused by carrier scattering as an effective way to boost the TE *PF*.

The paper is organized as follows: In Section II we present our methodology. In Section III we present our results for general model systems and elaborate on the electronic structure criteria that enable carrier filtering via inter-band scattering. In Section IV we present further results and discussions for realistic material considerations.

## II. METHODOLOGY

Our basis electronic structure consists of two overlapping bands as shown in Fig. 1a-b. In the first case the bands are of opposite type, i.e. we have a light conduction band and a heavy valence band to allow for asymmetry in the electronic structure. In the second case, we also examine the possibility of two bands of the same type (two conduction bands). Unless otherwise stated, we consider both intra- and inter-band scattering, i.e. we allow scattering within a certain band and from one band to the other, respectively. In Fig. 1c-d, we show the PF of these bandstructure materials versus the position of the Fermi level, $E_F$, with respect to the light CB edge, $\eta_F$, as a brief initial illustration. The black lines show the PF that is obtained by the single (light) band material, whereas the colored lines are the ones obtained by the overlapping band materials. A much larger PF peak is obtained when the Fermi level is placed well into the CB, at the point where the extremum of the heavy band resides, in regions that can be considered 'metallic'. The large PF is a consequence of selective scattering of the carriers with energy below the Fermi level on the heavy band as we explain below.

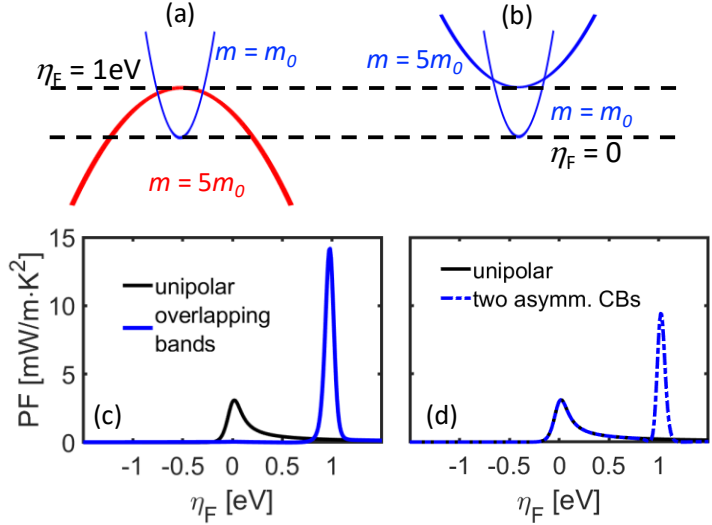

**Figure 1:** (a-b) Schematics of typical bandstructures investigated in this work. In (a) two asymmetric overlapping bands, a CB and a VB, are considered. This bandstructure mimics an *sd*-alloy with filled *d*-levels. In (b), two asymmetric CBs are considered. This bandstructure mimics an *sd*-alloy with empty *d*-levels. $\eta_F$ is the distance of the Fermi level $E_F$ from the light CB edge. Thus, $\eta_F = 0$ corresponds to the light CB edge, and $\eta_F = 1$eV indicates the upper/lower boundary of the overlapping region. (c) The PF for the bandstructure in (a) as a function of the position of the Fermi level (blue line), compared to the case of the stand-alone light CB (unipolar transport). Elastic phonon-scattering is considered. (d) The PF of the bandstructure in (b), with (dashed-dot) and without (solid) the heavy band. When the bands overlap, the scattering is always both *intra*- and *inter*-band. $T = 300$ K is assumed. The band parameters are $m_{CB} = 1\ m_0$, $m_{VB} = 5\ m_0$ (where $m_0$ is the electron rest mass), and scattering specific parameters are $D_{ADP} = 5$ eV, $\rho = 3$ g/cm$^3$ and $v_s = 6$ km/s. The scattering rates for these systems are shown in the Supplemental Material [51].

We compute electronic transport using the linearized Boltzmann transport equation (BTE) within the relaxation time approximation for parabolic bands. All simulations are performed at room temperature, $T = 300$ K. Within the BTE, the TE coefficients are given by [52]:

$$\sigma_{ij(E_F,T)} = q_0^2 \int_E \Xi_{ij}(E)\left(-\frac{\partial f_0}{\partial E}\right) dE, \tag{1}$$

$$S_{ij(E_F,T)} = \frac{q_0 k_B}{\sigma_{ij}} \int_E \Xi_{ij}(E)\left(-\frac{\partial f_0}{\partial E}\right) \frac{E-E_F}{k_B T} dE, \tag{2}$$

$$PF = \sigma S^2 \tag{3}$$

where $\Xi_{ij}(E)$ is the Transport Distribution Function (TDF) defined below in Eq. (4), $E_F$, $T$, $q_0$, $k_B$, and $f_0$, are the Fermi level, absolute temperature, electronic charge, Boltzmann constant, and equilibrium Fermi distribution, respectively. The TDF is given by:

$$\Xi_{(E,T)} = \frac{1}{3} v_{(E)}^2 \tau_{(E)} g_{(E)} \tag{4}$$

where $v_{(E)} = \sqrt{\frac{E}{2m_c}}$ is the parabolic band velocity involving the conductivity effective mass $m_C$, $g_{(E)}$ is the density of states (DOS), involving the $m_{DOS}$, and $\tau(E)$ is the relaxation time, which involves the total DOS. The factor $\frac{1}{3}$ picks the unidirectional velocity square component since the term $v^2$ combines partial velocities of all three Cartesian orientations. The use of parabolic bands in this work allows us to consider this simple form of the TDF without loss of generality in our conclusions. For simplicity, unless otherwise specified, we only consider elastic scattering with acoustic phonons within the deformation potential approximation (ADP), which for temperatures much larger than the Bloch-Gruneisen temperature (see below for further discussion) has scattering rates given by [53]:

$$\frac{1}{\tau_{nm(E)}^{(ADP)}} = \frac{\pi}{\hbar} D_{nm}^2 \frac{k_B T}{\rho v_S^2} g_{s,m(E)} \tag{5}$$

where $D_{nm}$ is the associated deformation potential which defines the scattering strength of the specific process, $\rho$ is the mass density, $v_s$ the sound velocity, and $g_{s,m(E)}$ is the scattering density of states (the DOS available for carriers to scatter into). The subscripts $n$ and $m$ correspond to the different bands that interract through the scattering process. We consider both intra-/inter-band transitions, which can be facilitated by different deformation potentials. In the intra-band scattering case, $m = n$ and we result to $\tau_{nn}$ with $D_{nn}$ (which we will refer to as $D_{intra}$), with the density of states $g_{s,n(E)}$ being that of the band on which the initial state resides. In the inter-band scattering case, $m \neq n$, which results to $\tau_{nm}$ with the associated $D_{nm}$ (which we will refer to as $D_{inter}$), with the density of states $g_{s,m(E)}$ being that of the other band.

In the last section of the paper (Section IV), to related more to realistic materials, we also address non-polar optical phonons in the framework of the optical deformation potential theory (ODP), for which the scattering rate is given by [53]:

$$\frac{1}{\tau_{nm(E)}^{(ODP)}} = \frac{\pi D_{nm}^2}{2\rho\omega_{\text{ph}}}\left(N_\omega + \frac{1}{2} \mp \frac{1}{2}\right)g_{s,m}(E \pm \hbar\omega_{\text{ph}}) \qquad (6)$$

where $g_{s,m}$ is the scattering density of states (that carriers scatter into) for absorption and emission, $N_\omega$ is the Bose-Einstein distribution function, $\omega_{\text{ph}}$ is the frequency of the phonons, and $D_{nm}$ is the inelastic optical deformation potential that defines the optical phonon scattering strength. The equation above accounts for both phonon emission and absorption scattering processes. Again, the subscripts *n* and *m* correspond to the different bands that interract through the scatering process.

The overall scattering rate from an initial state at energy *E* and band/valley *n*, combines scatteign into band/valley *n* and *m*, and is accounted using Matthiessen's rule as:

$$\frac{1}{\tau_{n,total}(E)} = \frac{1}{\tau_{nn}(E)} + \frac{1}{\tau_{nm}(E)} \qquad (7)$$

Note that in the case of alloys, alloy scattering will be the dominant mechanism, however, both ADP and alloy scattering are proportional to the density of states, thus considering only ADP comes without loss in generality. Also note it could be possible to derive analytical expressions for this model using such scattering simplicity for easy comparison to experiments, but the code we have implemented can be executed in seconds to fit this purpose as well.

The values for the deformation potentials we use, which define the scattering strength and the intra-/inter-band/valley scattering strengths, are mostly $D = 5$ eV and up to $D = 10$ eV for acoustic phonon scattering and $D = 10$ eV/Å for optical phonon scattering. These are chosen to be in the range of many prominent TE materials, semiconductors, and semimetals, as reported for various cases in the literature [54, 55, 56, 57, 58]. Typical semiconductors such as Si, Ge, GaAs also have deformation potentials in the range we use for both ADP and ODP [53, 59]. With regards to the values of the effective masses, we use masses of $1m_0$ for the light band and up to 10 $m_0$ for the heavy band which provides 'filtering-by-scattering', where $m_0$ is the rest mass of the electron. Indeed, effective masses of metals and alloys generally range in the region between 1-3 $m_0$ [60, 61, 62]. These values can be further enhanced by increasing valley degeneracy which increases the effective $m_{DOS}$ up to 11 $m_0$ [63], but can be even further increased by introducing flat bands to values even up to 30 $m_0$ [64]. Overall, the parameters we use such as deformation potential values and effective masses are similar to what is encountered in realistic systems. Our purpose in the paper is to focus on the physics of 'filtering-by-scattering' rather than the specific numbers, and the exact choice of parameters would not cause any quantitative changes to our arguments and conclusions.

Also note that our scattering parameters are chosen to provide PFs of ~ 3 mW/mK$^2$ for the unipolar case, typical for good TE materials, thus, the PF amplitudes achieved in the discussions throughout the paper can be quantitatively realistic. In general, the only parameters we employ are the effective masses of the material, and we then adjust the acoustic phonon scattering deformation potential to $D$ = 5 eV to obtain the PF for the unipolar case. The choice of the PF is in line with the PFs of good TE materials, that we use as a starting point. For example, in some of the prominent TE materials such as $Bi_2Te_3$, $Mg_3Sb_2$, PbTe, half-Heusler alloys and even Heusler semi-metals, etc., the PF (at various temperatures) ranges from 1-5 mW/mK$^2$ [59, 65, 66, 67, 68, 69, 70, 71]. This is the starting point, and we seek directions to increase it from here on. The PF values of metals are smaller due to their low Seebeck coefficient, typically having values below 1 mW/mK$^2$ [72, 73], but we have chosen a starting point of PF ~ 3mW/mK$^2$ since it reflects the PF of some high performing metals such as NiCu [74, 75, 76].

All our simulations are performed at room temperature, $T$ = 300 K. At that temperature we can safely assume that all phonon modes are excited for the majority of materials of interest, and isotropic scattering using the required phonon wavevectors that facilitate these transitions, is possible. For this, typically the operating temperature needs to be larger than the Bloch-Gruneisen temperature, which is the temperature above which phonons with momentum larger than $k_F$ are populated. Below that temperature, at $E_F$ only small angle scattering can be facilitated, since only small wavevector phonons are excited. The Bloch-Gruneisen temperature is given by $T_{BG} = \hbar v_s k_F/k_B$, where $\hbar$ is the reduced Planck's constant, $k_B$ is the Boltzmann constant, $v_s$ is the sound velocity of the material, and $k_F$ is the Fermi wavevector. The Bloch-Gruneisen temperature is less than 300 K for most materials, although the 300 K we consider is typically smaller that the Debye temperature, $\Theta_D$ (for example in common metals) [77, 78]. Typically, the Bloch-Gruneisen temperature is lower than $\Theta_D$, and it takes values around $\Theta_D/5$ up to $\Theta_D/2$ [62, 79]. In our specific examples, our Fermi energy of interest, $E_F$, which provides the PF peak, is around the overlap region between the light band and the heavy band extremum (and into the lighter band), which we set at 1eV in most simulations, and up to a maximum of 1.5 eV in one limiting case. By considering some relevant numbers, e.g. $v_s$=5000 m/s, and a Fermi wavevector up to even 0.3 of the Brillouin zone (BZ) length (which is already large), i.e. $k_F$ = 0.3 x $2\pi/a_0$, where we let $a_0$ = 0.5 nm, it turns out that $T_{BG}$ = 144 K, which is below half of our simulation temperature. In this scenario, to have $T_{BG}$ =300 K, $k_F$ would be as large as 60% of the BZ, which means that for the band of interest with $m^*$ = 1 $m_0$, which is responsible for the large PFs, the Fermi energy ($E_F = \hbar^2 k_F^2/2m^*$) would be $E_F > 2$ eV. Thus, we safely operate at a temperature above which phonons with momentum larger than $k_F$ are populated, and all carriers from the light band at the critical energy can scatter isotropically in the entire DOS of both the light and the heavy band (intra- and inter-band scattering). On the other hand, in the energy regions well into the heavy band, for full isotropic scattering (beyond small angle scattering with small phonon wavevectors), $k_F$ is large and large phonon wavevectors are necessary, essentially the Bloch-Gruneisen temperature increases as $E_F$ moves deep into that energy region. Although phonons with such large wavevectors will be certainly excited at 300 K, these phonons can have a substantial energy associated with them, or can reside on the optical branches as well, having energy $\hbar\omega$. In that case, the transitions will be inelastic and will be described better by Eq. 6 for optical/inelastic phonon scattering. To keep the scattering treatment simple and focus on the illustration of the high PF principle we present, in most of the paper we used only elastic/acoustic phonon scattering considerations. In Section IV we investigate the influence of optical/inelastic phonon scattering, but essentially reach the same conclusions. This is because any such scattering details do not affect the energy region of interest around the heavy band extrema, where the Fermi surface is small.

Also note that we chose to use electron-phonon scattering alone, since our study is at room temperature ($T$ = 300 K) and our aim is to illustrate a large PF design principle that is triggered by inter-band scattering. The specific details of the scattering mechanism involved are not of importance. To keep this simple, we do not consider for example mechanisms such as alloy scattering, or electron-electron scattering. In fact, any additional scattering will potentially increase the amplitude of the effect we describe. Specifically for electron-electron scattering, it is a mechanism that can be strong at low temperatures [62, 79], whereas at room temperature electron-phonon scattering typically dominates. In fact, computational studies in the case of bulk Si, have demonstrated that electron-electron scattering is not significant, even at high doping conditions [80, 81]. Its effect only being noticeable on the high energy tail of the distribution function, at energies of little relevance to our studies [81]. Even experimental data on the temperature dependence of the electrical conductivity in half-Heuslers, either metallic or semiconducting, have not reported the characteristic $T^2$ trend of electron-electron scattering at 300 K. Thus, this mechanism will not have an effect on our studies, either quantitative or qualitative, and we have not included it.

III.   MODEL SYSTEM RESULTS

We begin our discussion by examining the TDF as a function of energy, which turns out to be the dominant factor that determines the TE coefficients. Throughout the rest of the paper, we consider a mass of 1 $m_0$ for the light band and 5 $m_0$ for the heavy band in order to be closer to observed band structures in real materials such as *sd*-metal alloys and half-Heusler compounds (for the density of states). In Fig. 2a we present four simulation cases corresponding to the bandstructures shown in the sub-figures Fig. 2b-d below. The first bandstructure case we consider is the typical symmetric CB/VB wide bandgap (unipolar) material case, as shown in Fig. 2b, which we will use as a reference. The TDF for this case is shown by the red line in Fig. 2a. As expected, both in the CB and VB the TDF follows a linear trend with energy (since $\tau \sim 1/g$, then the energy dependence follows that of $v^2$, which is linear in *E*) [40]. Here we considered a bandgap of $E_g = 1$ eV. The second case we consider is the overlapping CB/VB case, but we still keep the two bands symmetric (Fig. 2c for bands, and black line for the TDF in Fig. 2a). Here, the bandgap is $E_g = -1$ eV, i.e. there is a -1 eV CB/VB overlap. The TDF acquires an upward shift outside the band overlap region, since the VB now begins at +0.5 eV and the CB at -0.5 eV. The overlap region has a finite TDF value as well, indicating a conducting energy region. Note that we allow for inter-band scattering in this region as well.

We now proceed in the asymmetric band case (Fig. 2d), for which the TDF is shown by the blue line in Fig. 2a. In this case we have increased the effective mass of the VB to $m_{VB} = 5\ m_0$. On the very left of the energy scale, the TDF slope and value drop compared to the overlapping symmetric case, a consequence of the heavier mass introducing more scattering and lower carrier velocities. In the far-right energy region, the TDF follows the overlapping symmetric case since the CB mass is the same in both cases ($m_{CB} = 1\ m_0$). The most interesting observation, however, is around energy $E = 0.5$ eV, in the region where the overlap between the VB and CB ends, towards the light mass CB. The TDF acquires a very sharp slope, which begins slightly below the end of the overlap region, and merges with the TDF of the non-overlapping bandstructure (black line). Note that the TDF in the overlap region is now strongly suppressed. This sharp slope allows for a finite Seebeck coefficient as we show below, despite the fact that we are considering the bandstructure of a metal, which typically have very low Seebeck values. As we show later on, this leads to very high PFs. The feature that enables this sharp slope is inter-band scattering of the light carriers of the CB into the heavy VB. At the energy point where the VB ends (or where the overlap ends), this dominant scattering mechanism disappears, and the carriers recover their high TDF. Indeed, the magenta line in Fig. 2a shows the case where we remove the inter-band scattering (i.e. from one band to the other), allowing only for intra-band scattering (i.e. within the same band). The sharp slope disappears and the overall TDF is now a superposition of the separate portions due to the CB and VB, and consequently the large PF performance will disappear as well, as we show later on. Interestingly, introducing excess scattering can be considered as an unconventional direction for TEs, because it reduces the electrical mobility and conductivity. However, it creates a scattering-induced energy filtering effect, which eliminates transport below certain energies, and consequently improves the Seebeck coefficient, in a similar manner to approaches such as energy filtering from potential barriers [31].

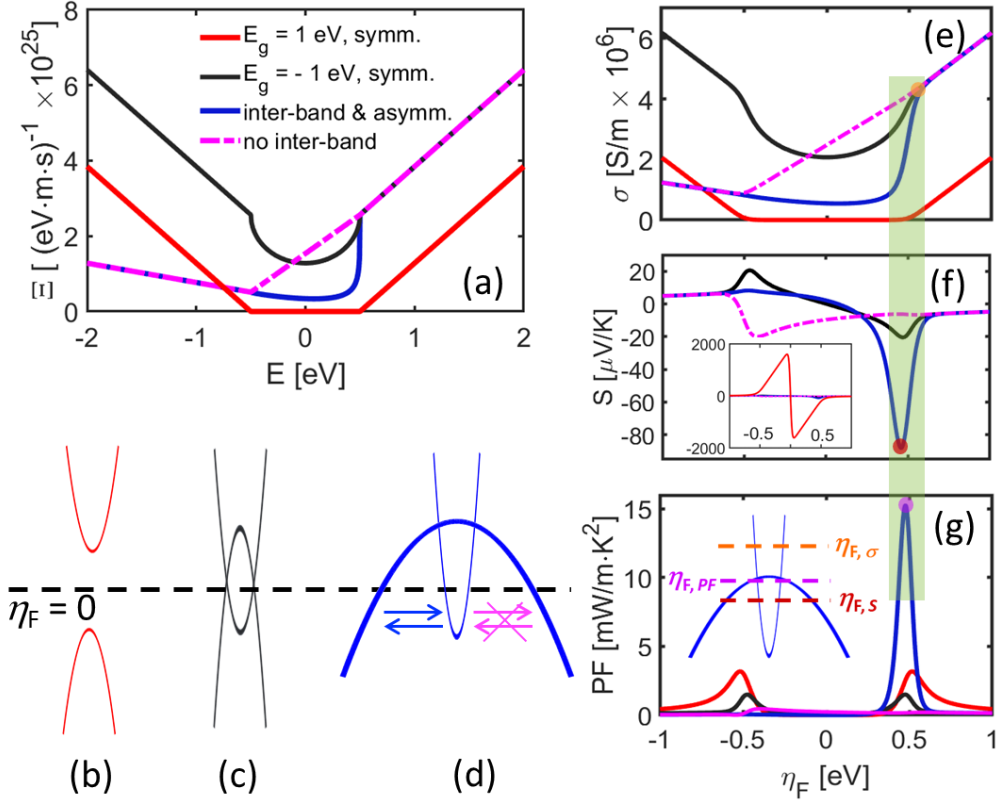

**Figure 2**: (a) The transport distribution function (TDF) for different bandstructures and scattering considerations as indicated in (b-d) using the same color coding. The relevant parameters for the band and transport simulations are $m_{CB} = 1\ m_0$, $m_{VB} = 5\ m_0$ (where $m_0$ is the electron rest mass), ADP-limited transport; $D_{ADP} = 5$ eV, $\rho = 3$ g/cm$^3$ and $v_s = 6$ km/s. The bandgap $E_g$ is as indicated, where $E_g < 0$ eV indicates band overlap. $E = 0$ eV corresponds to the middle of the overlapping region (or middle of bandgap). From the bandstructures considered, (b) is for symmetric bands with large bandgap, (c) for symmetric bands with overlap (negative bandgap), while (d) introduces asymmetry in the overlapping bands. For the latter, we consider the cases with (blue) and without (magenta) inter-band scattering. Sharp features are introduced in the TDF when both asymmetric bands and inter-band scattering are present. (e) Electrical conductivity, (f) Seebeck coefficient (the red line for the large bandgap case is shown in the inset), and (g) power factor, for the four cases shown in (b-c). $\eta_F = 0$ corresponds to the middle of the overlapping region (or middle of bandgap).
At the optimal position of the Fermi level, simultaneous increase of $\sigma$ and $S$ is observed as highlighted in green. The inset of (g) depicts the relevant position of the Fermi level in the case of asymmetric overlapping bands with inter-band scattering, for the highest $S$ ($\eta_{F,S}$), when $\sigma$ recovers the value of the light band alone ($\eta_{F,\sigma}$), and for the maximum PF ($\eta_{F,PF}$). These are shown by dots of correspondingly the same colors in panels (e-g).

The TE coefficients for the bandstructure cases of Fig. 2b-d are plotted in Fig. 2e-g (the line coloring in the two figures is the same). The electrical conductivity in Fig. 2e clearly follows the TDF of Fig. 2a, as the conductivity is a direct integration over the TDF.

The Seebeck coefficient for the sizable bandgap material follows the typical increase away from the band edges (+/- 0.5 eV) and into the gap, before the left and right branches collapse at midgap (which we now take as the reference 0 eV) and meet at zero as shown in the inset of Fig. 2f. In Fig. 2f the black and blue lines are for the cases of symmetric and asymmetric band overlap, respectively. As expected from a semi-metal, the Seebeck coefficient is very small in absolute terms, but importantly it is not zero. In fact, positioning the Fermi level around the region of the steep TDF (and conductivity), i.e. where the VB extremum appears, the Seebeck shows a noticeable increase. This is observed in the symmetric case (black line), but more strongly in the asymmetric band case (blue line). The increase in the Seebeck coefficient in this particular $E_F$ region, enables very high PFs, as shown in Fig. 2g. Compared to the PF of the large bandgap case (essentially unipolar transport), it is at least 5× higher. The symmetric overlapping band case (black) still has a significant PF for a semi-metal, but as expected, it is below the unipolar case (red line). In the case where the two bands do not scatter into each other (magenta line), the PF is also diminished. Thus, it is quite interesting that enhanced scattering (which is typically a drawback as it reduces the conductivity), and band overlap (typically avoided to reduce bipolar effects), can actually provide very large PFs. This is done by taking into advantage the large conductivity of the higher energy states of the light band, while acquiring a small Seebeck coefficient through 'filtering-by-scattering' of the lower energy carriers into the heavy band.

Importantly, we notice that when the Fermi level is positioned at the edge of the heavy band when the overlap ends, i.e. at $\eta_F = 0.5$ eV in Fig. 2e-g, a *simultaneous* increase in both $\sigma$ and $S$ is achieved (highlighted regions). This is the basic reason that the PFs can reach such incredibly large values. It is a direct consequence of the sharp feature created in the TDF around those energy values, allowing for energy asymmetric conduction (i.e. substantial difference in transport in the energy regions below/above the $E_F$). An alternative view is that this increases the energy of the current flow, which is the thermodynamic definition of the Seebeck coefficient [82,83]. To the left/right of the sharp feature in the TDF and conductivity, $S$ gradually reduces back closer to zero, as the slope of the TDF decreases, generating a peak in $S$ around the sharp TDF step. On the other hand, by placing the Fermi level near the energy of the sudden step in the TDF, the $\sigma$ sharply increases as the strong inter-band scattering ends and the states of the light band at those energies have high velocity and conductivity. Note also that the Fermi level position of maximum $S$ appears at $\eta_{F,S} = 0.46$ eV, still into the heavy band, at 40 meV below its band edge. This is lower than the point when $\sigma$ reaches the values characteristic of the light band ($\eta_{F,\sigma} = 0.61$ eV, 110 meV above the heavy band edge). This is sketched in the inset of Fig. 2g, where $\eta_{F,S}$, $\eta_{F,\sigma}$, and the Fermi level position of maximum PF, $\eta_{F,PF}$, are shown, with the latter appearing very close to the VB extremum.

Thus, to realize the effect that we describe, we need both, inter-band scattering and band/transport asymmetry, in addition to band overlap. In Fig. 3 we examine these three essential ingredients one-by-one by simulating the TE coefficients, electrical conductivity, Seebeck coefficient and PF, and by varying: i) the degree of inter-band scattering (first column, Fig. 3a-c), ii) the degree of band asymmetry (second column Fig. 3d-f), and iii) the degree of band overlap (third column, Fig. 3g-i). In all cases, we show by the blue line our reference case as presented earlier in Fig. 2e-g.

In the first column, we change the degree of inter-band scattering by varying the deformation potential that we use to form the inter-valley acoustic phonon scattering rates ($D_{\text{inter}}$). In black we show the case where no inter-band scattering is considered, while we consider $m_{VB} = 5\ m_0$ (versus $m_{CB} = 1\ m_0$) and $E_g = -1$ eV of overlap. When we increase the inter-band scattering strength, the electrical conductivity drops overall. However, the envisioned sharp conductivity feature does appear, and then becomes more defined as we increase inter-band scattering in the case shown by the green line (Fig. 3a). Similarly, the Seebeck coefficient in Fig. 3b increases around that sharp conductivity $\eta_F$ region, which follows the TDF steepness (shown in the inset). This drives the PF to very high values as well (Fig. 3c). In this column we also plot the extreme case where the TDF is completely vertical at the end of the overlapping region and zero below the VB edge; this corresponds to the maximum theoretical limit of the PF rise. We do this by artificially 'cutting' the TDF below the overlap edge, enforcing zero conductivity from the valence band edge energy and below. This is an unphysical scenario, but it provides the upper limit of the PF that can be achieved. This situation is shown by the blue dotted lines in Figs. 3a-c. Ultimately, PFs up to ~ 40 mW/mK$^2$ can be reached. Note that still in our case, the light band materials is the band of a semiconductor. A metal or a lighter semiconductor with much higher conductivity would allow for much larger PFs.

Note that we have arbitrarily changed the deformation potentials of the inter-band process to values even higher than the intra-band process to consider much faster inter-band scattering rates compared to intra-band scattering rates in general. Some of the strong scattering mechanisms in thermoelectric materials such as polar optical phonon scattering and ionized impurity scattering are anisotropic processes and favor small momentum exchange vectors ($q$-vectors), which makes them intra-valley. However, they can have an equally strong inter-valley scattering component if the valleys of the final bands are in close proximity in k-space with the initial band, i.e. if they are located at the same high symmetry point for example. There are also other physical scattering processes for which the inter-valley processes are stronger than the intra-valley processes. For example, for half-Heusler materials the inter-valley optical phonon processes (governed by $D_{ODP}$) are much stronger compared to the intra-valley acoustic phonon processes (governed by $D_{ADP}$) as discussed by Zhou et al. [55]. In common semiconductors like p-type Si (where all valleys are located at the $\Gamma$-point), ab initio simulations have shown that ODP processes are stronger compared to ADP processes, especially for energies beyond the onset of phonon emission [53, 59]. In the cases we consider, the elongated semi-flat bands encountered in metals (formed by $d$- or $f$-orbitals), can span large portions of the k-space region and can have overlap with the dispersive $s$-orbital bands in k-space. In this case the short $q$-vector scattering rates between intra- and inter-valley transitions are of similar strength. Additional long-range $q$-vector processes such as ODP could make inter-valley scattering stronger compared to intra-valley scattering.

In any case, the intention here is to consider stronger inter- compared to intra-band scattering rates, which is also captured by elongated final bands, or final bands with larger effective masses, as examined in the second column of Fig. 3. In this case, our starting point is an overlapping band with $E_g = -1$ eV with inter-band scattering considerations using $D_{inter} = 5$ eV. Here we vary the mass asymmetry around our reference $m_{VB} = 5m_0$ by considering $m_{VB}$ of $2m_0$ (black line) and $10m_0$ (green line). Again, the larger the asymmetry, the sharper the conductivity feature is around the VB extremum (Fig. 3d), the larger the Seebeck coefficient (Fig. 3e), and the larger the PF (Fig. 3f).

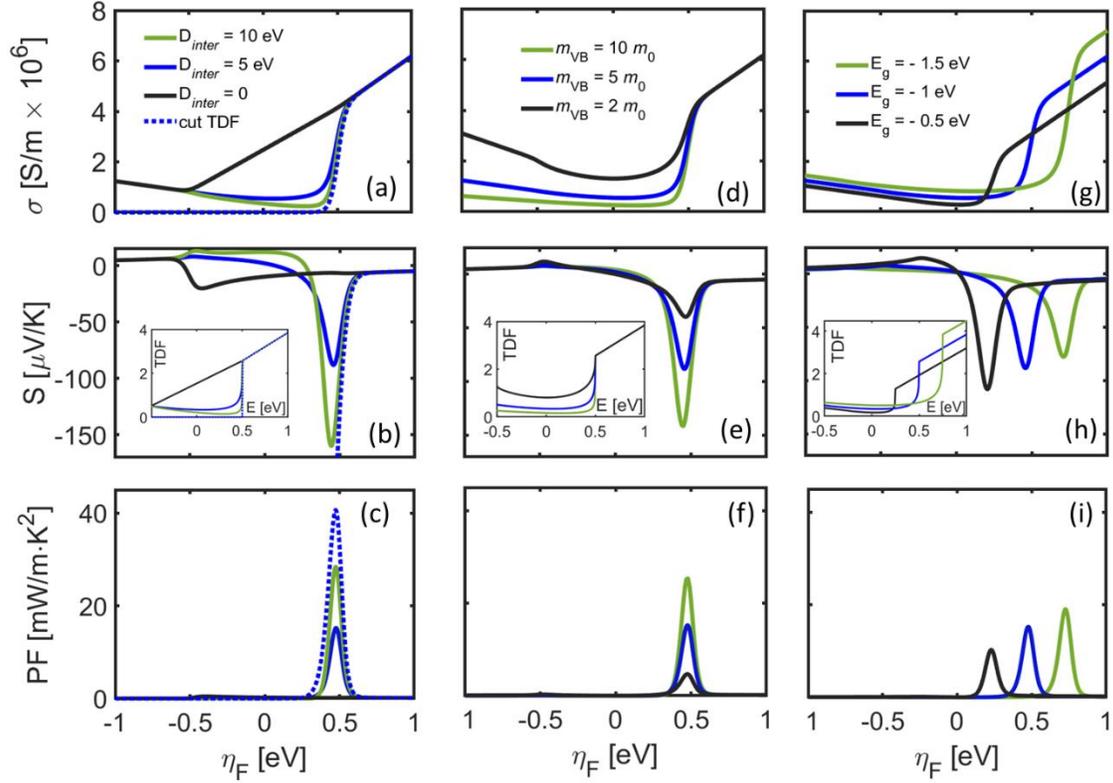

**Figure 3**: Investigation of the parameters that create the sharp features in the TDF and lead to high PFs in the bandstructure composed of the light CB and heavy VB as shown in Fig. 2d. The reference case is the one shown in Fig. 2 by the blue line (as well as here). The first row shows the electrical conductivity, the second the Seebeck coefficient and the third the power factor. (a-c) First column: The role of *inter-band* scattering by varying the *inter-band* scattering deformation potential $D_{inter}$ around the reference. Here, the black line indicates no inter-band scattering for comparison (provides miniscule PF). The dashed-blue line indicates the limiting case of zero conductivity from the edge of the VB energy and below, achieved by setting the TDF to zero in that region ('cut TDF'). This will provide the theoretical upper PF limit for this effect. (d-f) Second column: The role of band asymmetry, where the upward CB is kept fixed, while the effective mass of the downward VB-like band varies around the reference. (g-i) Third column: The role of the CB-VB energy overlap, for different degrees of overlap. The insets in (b), (e) and (h), show the TDF, $\Xi$, in units of $10^{25}/(eV \cdot m \cdot s)$ around the steep feature. Overall, tuning of *inter*-band scattering and asymmetry controls the Seebeck coefficient, while the overlap tuning controls the conductivity more. $\eta_F = 0$ indicates the middle of the overlapping region.

Finally, in the third column, we change the degree of band overlap from $E_g = -0.5$ eV (black line) to $\underline{E}_g = -1.5$ eV (green line). Our basis is again the asymmetric band with $m_{VB} = 5\ m_0$, $m_{CB} = 1\ m_0$, with inter-band scattering considerations using $D_{inter} = 5$ eV. Sharp features in $\sigma$ are also created as the overlap increases (Fig. 3g). On the other hand, contrary to the previous two cases, the Seebeck coefficient decreases (Fig. 3h). This is because by increasing the overlap between the heavy and the light bands, the step in the TDF becomes smoother; for example, for the 0.5 eV overlap (black line) the TDF increases by 10× in an energy window of 0.2 eV whereas for the 1.5 eV overlap (green) the TDF increases by 7× in a carrier energy window of 0.4 eV. The PF, however, overall increases with band overlap. Thus, each of these three criteria (inter-band scattering, band/ transport asymmetry and band overlap) is essential for large PFs for this design concept. Out of the three, we note that asymmetry can be beneficial in different settings as well. For example, we have shown in Ref. [14, 24] that small bandgap semiconductors with asymmetric bands could allow for large PFs. Interestingly, asymmetry and inter-band scattering increase the PF by increasing the Seebeck coefficient, whereas increasing the band overlap enhances the PF by improving the conductivity.

We now proceed in Fig. 4 to examine the design parameters for the case of two overlapping CBs. A schematic of these bands is shown in Fig. 1b. In this case, the reverse happens, where transport in the light CB is favored below the extremum of the heavy band, rather than above. Here, the TDF drops at and into the heavy band edge. While the details of the $\sigma$ and $S$ behavior are reversed in energy, conceptually the TDF steepness variation results in a PF improvement. As a consequence of the steep feature, the positive impact on the PF of the inter-band scattering strength, the asymmetry, and the overlap, is maintained. Notable differences are that the Seebeck coefficient changes sign, enabling to achieve positive $S$ values in metals, and that an additional peak appears at the edge of the light band. The latter reflects the semiconducting character of the bandstructure when the Fermi level is placed in that energy range, and is well-described by unipolar transport characteristics. Much like Fig. 3, we observe that the inter-band scattering and the band mass anisotropy improve the PF via the Seebeck coefficient, while the degree of overlap boosts performance by increasing $\sigma$.

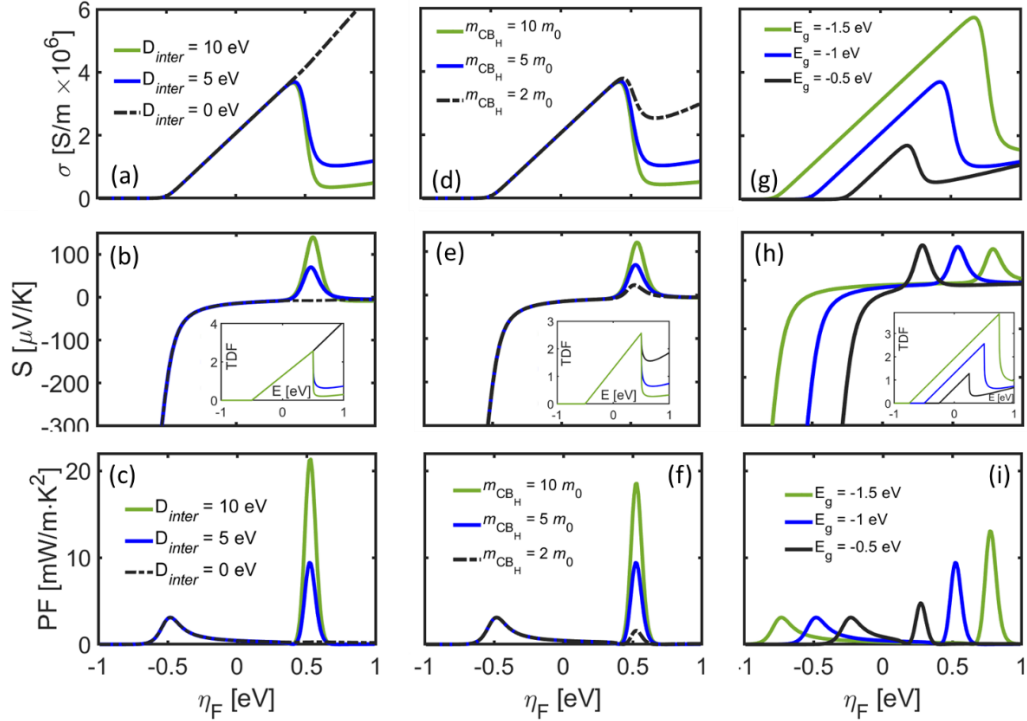

**Figure 4**: Investigation of the parameters that create the sharp features in the TDF and lead to high PFs in the bandstructure composed of two overlapping CBs, a light and a heavy one as shown in Fig. 1b. The reference case is the one shown by the blue lines. The first row shows the electrical conductivity, the second the Seebeck coefficient and the third the power factor. (a-c) First column: The role of *inter-band* scattering by varying the *inter-band* scattering deformation potential $D_{inter}$ around the reference (black line indicates no inter-band scattering). (d-f) Second column: The role of band asymmetry, where the lowest energy CB is kept fixed, while the effective mass of the highest energy CB varies around the reference line. (g-i) Third column: The role of the CB-CB band energy overlap, for different degrees of overlap. The insets in (b), (e) and (h), show the TDF, $\Xi$, in units of $10^{25}/(eV \cdot m \cdot s)$ around the steep feature. Overall, tuning of *inter*-band scattering and asymmetry controls the Seebeck coefficient, while the overlap tuning controls the conductivity more. $\eta_F = 0$ indicates the middle of the non-overlapping region, which is more intuitive to be considered as a common reference for all cases of overlapping bands. Note: this reference is different from what used in Fig. 1 where $\eta_F = 0$ was chosen at the band edge (since we had a case for unipolar transport there).

## IV. REALISTIC MATERIAL CONSIDERATIONS

Above we discussed how carrier 'filtering-by-scattering' can allow for ultra-sharp transitions in the TDF, which lead to *simultaneous* improvements in the electrical conductivity and Seebeck coefficient. This relaxes one crucial bottleneck of TE performance and allows the realization of very high PFs. In this section we discuss features that are encountered in realistic materials which could possess a suitable bandstructure. We discuss the presence of multiple valleys, the dominance of inelastic scattering mechanisms rather than elastic, and the presence of narrow impurity bands which overlap with the CB or VB, as well as elongated flat bands. Such effects, for example, can be traced among the transition metal alloys and the half/full-Heusler alloy families, which are materials with high TE potential, but also topological materials, 2D materials, and others [2,84,85]. In the Supplemental Material we present a list of possible materials (not an exhaustive list) which have large DOS variation and could possibly allow for large PFs [51]. We also provide an indication for the Seebeck coefficients in those materials, showing that sizable Seebeck coefficients, both n-type (negative) and p-type (positive), can be realized.

In the case of transition metal alloys, high PFs can be achieved when a very light (delocalized) *s*-band, and a very heavy (localized) *d*- or *f*-band, with large overlap and strong scattering between the *s*- and *d*-/*f*-bands is encountered, as recently experimentally demonstrated [33]. Ideally, the *s*-band should also have low intra-valley scattering rate to allow for a large mobility. Indeed, in reference [33] it was shown experimentally that alloys of the metal NiAu feature a very high PF when the alloy composition is such that the Fermi level is placed within the dispersive *s*-bands, but at the upper edge of the filled *d*-bands.

In the case of Heusler materials, full-Heuslers that are typically metals can be of consideration, but even semiconducting half-Heuslers (HHs) can be considered if alloying parental compounds with low bandgap can realize an overlap between the VB and the CB. One of the two bands should possess a light mass with large mobility and the other must have a very high $g(E)$. In all cases, inter-valley scattering between the CV and VB is required. In the case of Heuslers, the VB typically consists of many valleys. Note that our simulations show that intra-band scattering or inter-band scattering between the VB valleys does not affect the high PF, which is achieved in the light CB region of the bandstructure (in fact, inter-valley scattering in the VB can be marginally beneficial to the PF by enhancing the steepness of the TDF and the steepness of the conductivity around the onset of the sharp feature – we provide a brief explanation for this in the Supplemental Material [51]). Interestingly, this direction is the opposite of what is needed in the case of band alignment strategies to improve the PF, where strong inter-band scattering is detrimental [15,16,17,18,41,66]. Also note that although we performed all our analysis using a heavy VB and light CB, this 'filtering-by-scattering' approach in principle allows for both polarities. In the case of metal alloys, a *p*-like polarity can be obtained by placing the Fermi level below the edge of the localized band, likely to exploit the elements with $d/f$-levels. For half-Heuslers, the polarity is determined by the lighter of CB and VB and this can be tuned by the choice of the parental materials.

However, we stress that to tackle realistic materials, we must consider inelastic scattering with optical phonons, because they allow large momentum transfer, and hence they can facilitate the scattering process between bands, which reside far from each other in the Brillouin zone (BZ). For scattering into multiple valleys, the most effective scattering is from non-polar optical phonons. Polar optical phonon scattering (POP) is a strong mechanism for scattering into nearby valleys [57], but it is an anisotropic mechanism, thus its scattering rate decreases with the momentum exchange vector, weakening the filtering effect, but also are likely screened in metals. Thus, we address non-polar optical phonons in the framework of the optical deformation potential theory (ODP), as described by Eq. 6 above.

Specifically, we consider a bandstructure which mimics Heusler metals, we adopt one light CB, $m_{CB} = 1\ m_0$, and 12 VB valleys with an isotropic mass of $m_{VB} = 0.558\ m_0$ for each [40, 48], with negative bandgap, as depicted in Fig. 5a. The number of VB valleys reflects the valleys position in Γ, X (6-fold multiplicity, shared between two BZs, i.e. 3-fold degeneracy) and W (24-fold multiplicity, shared by three BZs, 8-fold degeneracy). This specific situation mimics the case of some metals with Heusler structure [85–91] as well as some other intermetallic compounds [87,88,92,93,94]. In Fig. 5a these are reflected to the central VB and two side VBs in each direction. The value of $m_{VB}$ is chosen to obtain a weighted DOS mass of 5 $m_0$. We consider arbitrarily a phonon energy of $\hbar\omega = 40$ meV, which, however, is similar to the optical phonon energies of many metals and semimetals (see table in Supplemental Material [51]). The inter-valley scattering deformation potential among the VBs is kept at $D_{inter} = 5$ eV/Å, while the inter-band deformation potential between the CB and the several VBs, $D_{CB/VB}$, is varied. We also consider intra-valley scattering with deformation potential of $D_{intra} = 5$ eV. In Fig. 5c-d we show the TE coefficients $\sigma$, $S$, and PF, respectively. The TDFs are shown in the inset of Fig. 5d. The blue dotted line shows the ultimate case, where the TDF is 'cut' below the valence band maxima (zero conductivity below those energies). The observations with respect to PF improvements are very similar to those observed earlier in the case of elastic scattering. The inset in Fig. 5e shows a schematic of the relevant positions of the Fermi level as in Fig. 2e-g, with the Fermi level position at maximum $S$, $\eta_{F,S}$, its position when $\sigma$ reaches the values characteristic of the light band, $\eta_{F,\sigma}$, and its position at maximum PF, $\eta_{F,PF}$. What is different from the case of elastic inter-band scattering, however, is that the position of $\eta_{F,S}$ and $\eta_{F,PF}$ nearly coincide, and they are placed above the heavy band edge, into the light band at an energy distance somewhat smaller than the phonon energy. This occurs when the dominant phonon emission scattering process from light band states into the heavy band weakens. The trend observed in Fig. 5b suggests that the position of the Fermi level must be tuned by taking into account the average optical phonon energy. Indeed, the optimal Fermi level position for the PF, as shown in Fig. 5b, varies linearly with phonon energy, such that the distance from the light CB band edge is somewhat smaller than $\hbar\omega$ (it appears that it is around a $k_B T$ smaller than $\hbar\omega$, but at this point we don't have further justification for this). This example shows that high PFs can also be achieved if the VB is composed of many (lighter) valleys, rather than just one heavy, provided that inter-valley scattering from CB to VB is strong. Note that we have extended our optical phonon energy range in Fig. 5b from 25 meV to 100 meV to cover a range of possible candidate materials. However, the 40 meV value we used for Fig. 5c-e is more close to the optical phonon energy of many metals and semimetals, and in particular $Fe_2VAl$ (45 meV) [95].

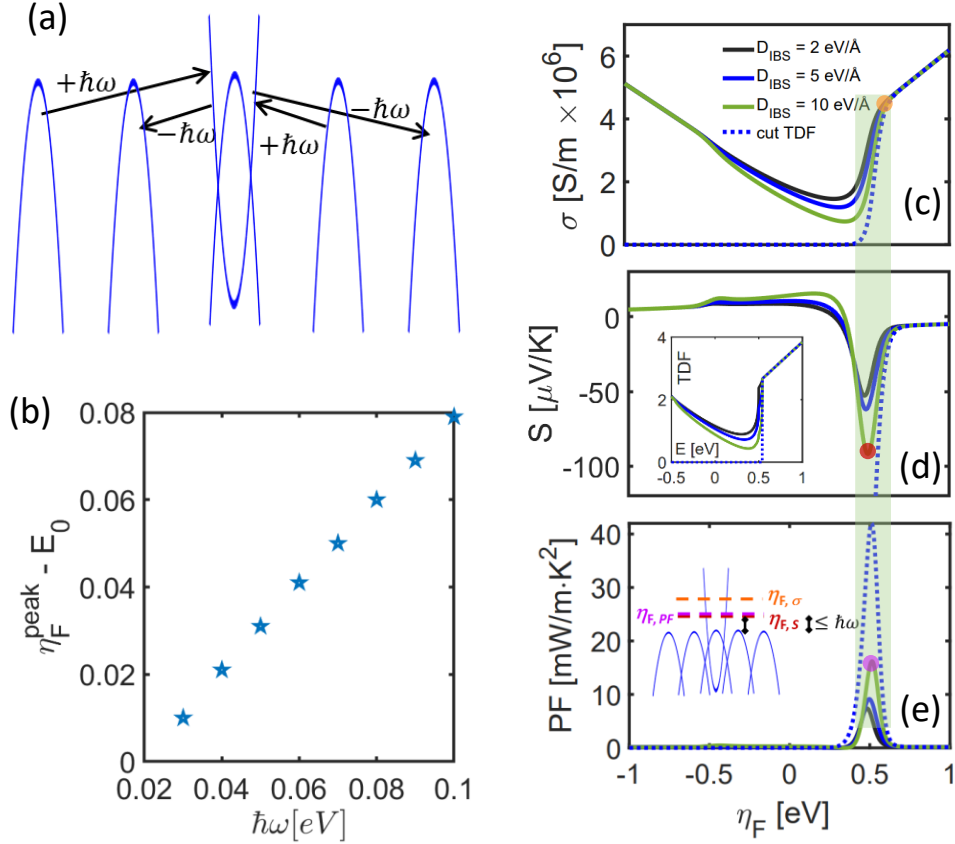

**Figure 5**: Mimicking the case of a HH-like metallic band structure, with a VB with maxima in Γ, X, and W. (a) Schematic of the bands and *inter*-band scattering transitions (IBS), where $\hbar\omega$ is the absorption/emission phonon energy. (b) The position of the Fermi level at which the PF peaks as a function of the phonon energy, where $E_0$ is the VB upper edge. The Fermi level zero reference, $\eta_F = 0$ eV, is taken to be at the middle of the overlapping region, i.e. $E = 0$ eV at the middle of the overlapping region as well. (c-e) The electrical conductivity $\sigma$, Seebeck coefficient $S$, and power factor, PF, for the specific case of $\hbar\omega = 40$ meV and different inter-band scattering strengths between the CB and the VBs. The blue dotted line shows the ultimate case, where the TDF is 'cut' below the valence band maxima (zero conductivity below those energies). The inset in (d) shows the TDFs for the different cases in units of $10^{25}/(\text{eV}\cdot\text{m}\cdot\text{s})$. The inset in (e) depicts the relevant position of the Fermi level for the highest $S$ ($\eta_{F,S}$), the point where $\sigma$ recovers the value of the light band alone ($\eta_{F,\sigma}$), and the point of maximum PF, ($\eta_{F,PF}$). These are shown by colored dots on the respective lines in panels (c-e). The scattering rates for this system is shown in the Supplemental Material [51].

Finally, to better mimic realistic metallic materials and alloys, with complex bandstructure features, we investigate electronic structures consisting of narrow bands overlapping with the common dispersive parabolic band. In this case, as shown in Fig. 6a, and 6e, we approximate the heavier overlapping band using two segments of intersecting parabolas, one upward pointing and the other downward pointing, each with a mass of 5 $m_0$. In comparison to the metallic materials of interest, as shown in the Supplemental Material [51] (taken from the Materials Project webpage), our fictitious bandstructures essentially fold the elongated realistic bands, but can also mimic electronic structures with impurity bands and alloy material electronic structures with broadened fractured/split bands [96]. For scattering and transport purposes, our simplistic band formation is a good first order approximation of realistic bands, but its implementation on the parabolic band model is much easier.

In Fig. 6a-d we consider the case of one narrow band of increasing width, and in Fig. 6e-h the case of increasing number of bands of the same width (with 'width' we define the extent of the band in energy). Figure 6a shows the DOS of the light band and of the three cases of a narrow overlapping band of width 0.1 eV (black), 0.2 eV (blue), and 0.3 eV (green). Here we choose to place these bands with an initial starting point to the left side and increase the height and width 'manually by construction' (which also increases its centered energy value). Schematics of the actual bands are provided in the Supplemental Material [51]. The corresponding TE coefficients are shown in Fig. 6b-d. We observe that a larger band width is more effective in filtering out the low energy states in the light band, boosting the TE performance. For the same reasons discussed so far, three PF peaks appear, one at the edge of the light band (conventional situation for a semiconductor TE material), and two peaks at the edges of the narrow band (which would be observable in band overlapping metals). The higher PF appears when the Fermi level is placed at the higher energy edge region of the wide band, which utilizes the higher velocity carriers and conductivity of the light band.

The effect of the number of narrow overlapping bands is described in Fig. 6e-h. Here we consider narrow bands of width 0.1 eV and increase their number, from one to three, at different positions close to the light band edge, as depicted in Fig. 6e. We start with the black colored band as our reference, and then we add the blue colored band and then the green colored band. The corresponding TE coefficients are shown in Fig. 6f-h. By increasing the number of bands, more filtering windows appear with corresponding oscillations in the conductivity, Seebeck and PF (and in the TDF), when plotted versus the Fermi level position. The relatively mild PF improvement can be understood by considering that the filtering efficiency, because of inter-band scattering, is proportional to the DOS of the heavy band. Thus, by increasing the number of identical bands, the only improvement is associated with the highest TDF value reached at the highest energy filtering window: the higher the energy of the top-most flat band, the faster the carriers in the light band, and the higher the PF peak. However, this will depend on the specifics of the bands, their amplitudes and their widths.

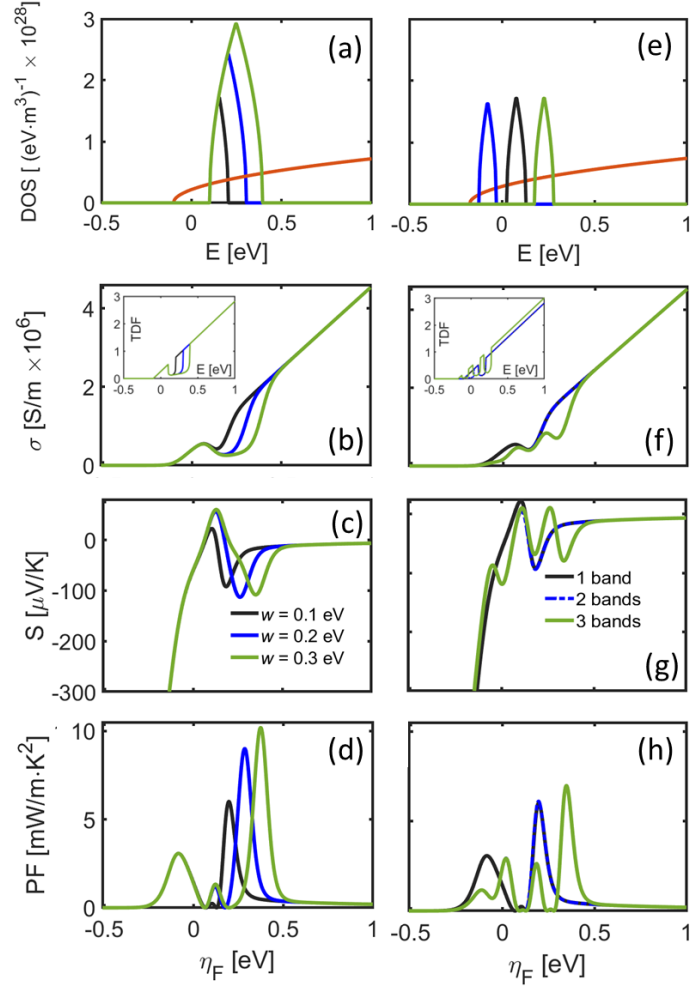

**Figure 6**: The case where heavy narrow bands overlap with a light CB. (a-d) Left column: The width $w$ (and height) of the narrow bands vary. (e-h) Right column: The number of the narrow bands varies. The first row shows the resulting DOS, the second row the $\sigma$, the third row the $S$, and the fourth row the PF. (a) The DOS for each narrow band we consider, on top of the light parabolic band in orange. In black, blue, green, the narrow bands have width $w$ of 0.1, 0.2, 0.3 eV, respectively (their extend in energy). (b-d) The corresponding transport coefficients. The inset of (b) shows the TDF in units of $10^{25}/(eV \cdot m \cdot s)$. A wider band enhances filtering-by-scattering and increases the PF. Three PF peaks can appear for each case, one at the edge of the light band, and two at the two edges of the narrow band. (e) The DOS for the case of one, two and three narrow bands of width 0.1 eV, placed incrementally next to each other into the light CB (i.e. we show the results for one narrow band, then two and then three). (f-h) The corresponding transport coefficients. Peaks appear in the PF at the position of the narrow bands' edges.

Thus, in this case of narrow bands, the filtering of low energy carriers in the CB, as enabled by the inter-band scattering, is retained. The most favorable situation occurs when the DOS of the narrow band is larger, and the upper edge is placed high enough in energy to reach the higher velocity carriers in the light dispersive band (and the Fermi level is also placed at those energies). Moreover, increasing the number of these narrow bands improves the TE performance, but the relative benefits are smaller compared to a single wider band with larger DOS.

Note that in alloy material engineering, in many cases improved Seebeck coefficients are observed in experiments, usually attributed to band convergence ($N_v$) increasing effective mass of materials (as extrapolated on the Pisarenko plot). Selective carrier filtering has been invoked as a route for increased Seebeck in nanostructured materials. However, in these situations, impurity bands, or broadened 'fractured' bands can appear, often evident in the unfolded bands from supercell DFT simulations [96]. The simulations in Fig. 6. mimic such "fractured" band effects, and show that indeed the appearance of such states in the band edges on their own can lead to large Seebeck coefficient variations (and oscillations with $E_F$) and PF improvements as well. Improvements in PF in alloyed materials therefore do not necessarily require band convergence, but can be a consequence of carrier filtering effects due to the emergence of multiple broken bands.

While details can vary, sharp features in the shape of quantities that determine electronic transport are suggested broadly to improve thermoelectric performance. The Mott formula indicates that sharp features in the DOS improve the Seebeck coefficient, which also triggered the work on low-dimensional thermoelectrics [97, 98], the work on resonant levels in the DOS [32], the works on energy filtering from potential barriers [31, 99], and lately works on flat bands in the bandstructure [12]. Mahan and Sofo showed that the highest $ZT$ (and efficiency) is provided by a Dirac δ-function transport distribution function (TDF) shape [100]. Such a TDF results in zero electronic thermal conductivity, because it enforces a zero variance to the energy dependence of the current flow. Further works showed that depending on the scattering and dispersion details, broadened δ-function TDFs can be more beneficial [101]. In general, however, the realization of sharp features is not the typical scenario in TE material design, and any new method that points towards that, as the one we describe, and especially of the degree we describe, would provide a novel direction for further exploration. In addition, having sharp features in the DOS that increase the thermopower, do not always translate to power factor improvements. They can cause reduction in the conductivity (adverse interdependence), which can lead to only moderate improvements, or even reductions in the power factor. In that respect, in this work, we show that sharp features in one side of the TDF, succeeded by a broadened increasing trend on the other side, increase the Seebeck coefficient and retain high conductivity. A simultaneous increasing trend is observed for both quantities in the relevant energy interval, leading to the increase of the power factor to ultra-high values. This TDF shape is very similar to what observed in special designs for energy filtering by the use of potential barriers [27, 99]. Note also, that in reality non-idealities and disorder will introduce band tails and smearing of the sharp edges that are created in the vicinity of the heavy band edge and are responsible for the large Seebeck coefficient increases. Such non-idealities will potentially reduce the sharpness of the TDF around that region, and in that case, the large increases in the Seebeck coefficient and power factor (PF) we observe, will be slightly mitigated (see Supplemental Material [51]).

Finally, we add a note on the *ZT* figure of merit. In general, metallic systems have high thermal conductivity, typically dominated by the electronic thermal conductivity part, which reduces their overall thermoelectric efficiency. In this scenario, however, the *ZT* will be determined solemnly by the Seebeck coefficient and the Lorenz number, *L*, as $ZT = S^2/L$. Thus, focus on boosting the Seebeck ceofficient is important. Additionally, the focus can also be on metallic systems with lower Lorenz number, by identifying materials in which the ratio of *L*, (which depends on the material's intrinsic properties) to the Lorenz constant $L_0 = \pi^2 k_B^2/(3q^2)$ (a fixed value) is less than 1, as observed in Mn (0.86), Ni (0.76), FeAl (0.90), and other bimetallic alloys at room temperature [102]. Even lower Lorenz numbers can be found, as for example for pure Al at low temperatures (<0.5) [103]. This adjustment allows for the thermal conductivity to be less dependent on the electrical conductivity. Consequently, further enhancing the electrical conductivity can improve the overall efficiency of the metallic system.

V. CONCLUSIONS

To conclude, we have investigated the thermoelectric power factor of metallic materials in which a light conduction band overlaps with a valence band of increased density of states, using a parabolic multi-band model and Boltzmann transport. By allowing inter-band scattering within electronic structures, we show that carrier filtering-by-scattering can be achieved, which drastically reduces the transport of carriers at the low energy region of the light conduction band. This situation leads to very sharp features in the transport distribution function, allowing metallic materials to acquire a sizable Seebeck coefficient due to the resultant built-in asymmetry in transport. In combination with the high conductivity of the remaining high energy states, ultra-high thermoelectric power factors, at levels of an order of magnitude (tens of mW m$^{-1}$ K$^{-2}$, i.e. even up to 40 mW m$^{-1}$ K$^{-2}$ in the examples we consider) compared to the state-of-the-art materials, can be achieved. In particular, we show that when the Fermi level is placed in the vicinity of the sharp features in the transport distribution function, the rare event of *simultaneous* improvement in the Seebeck coefficient and electrical conductivity can be realized. Our work will be useful for the design of thermoelectric materials with ultra-high power factors, unconventionally from the family of metals, a novel trend which resurfaces [33], but is yet largely unexplored.

## Acknowledgements

This work has received funding from the UK Research and Innovation fund (project reference EP/X02346X/1) and from the European Union–Next-Generation EU via the Italian call PRIN 2022, project code 2022XZ2ZM8.

References

[1] D. Beretta et al., *Thermoelectrics: From History, a Window to the Future*, Materials Science and Engineering R: Reports **138**, 210–255 (2019).

[2] C. Artini et al., *Roadmap on Thermoelectricity*, Nanotechnology **34**, 292001 (2023).

[3] G. K. Ren, S. Wang, Z. Zhou, X. Li, J. Yang, W. Zhang, Y. H. Lin, J. Yang, and C. W. Nan, *Complex Electronic Structure and Compositing Effect in High Performance Thermoelectric BiCuSeO*, Nat. Commun. **10**, 2814 (2019).

[4] Y. Zhu D. Wang, T. Hang, L. Hu, T. Ina, S. Zhan, B. Qin, H. Shi, L. Su, X. Gao, L. D. Zhao *Multiple Valence Bands Convergence and Strong Phonon Scattering Lead to High Thermoelectric Performance in P-Type PbSe*, Nat. Commun. **13**, 4179 (2022).

[5] J. Zhang and B. B. Iversen, *Fermi Surface Complexity, Effective Mass, and Conduction Band Alignment in n-Type Thermoelectric $Mg_3Sb_{2-x}Bi_x$ from First Principles Calculations*, J. Appl. Phys. **126**, 085104 (2019).

[6] Y. Pei, H. Wang, and G. J. Snyder, *Band Engineering of Thermoelectric Materials*, Adv. Mater. **24**, 6125 (2012).

[7] L. Wang, X. Zhang, and L. D. Zhao, *Evolving Strategies Toward Seebeck Coefficient Enhancement*, Acc. Mater. Res. **4**, 448 (2023).

[8] X. Wang, V. Askarpour, J. Maassen, and M. Lundstrom, *On the Calculation of Lorenz Numbers for Complex Thermoelectric Materials*, J. Appl. Phys. **123**, 055104 (2018).

[9] D. Parker, X. Chen, and D. J. Singh, *High Three-Dimensional Thermoelectric Performance from Low-Dimensional Bands*, Phys. Rev. Lett. **110**, 146610 (2013).

[10] J. P. Heremans, V. Jovovic, E. S. Toberer, A. Saramat, K. Kurosaki, A. Charoenphakdee, S. Yamanaka, and G. J. Snyder, *Enhancement of Thermoelectric Efficiency in PbTe by Distortion of the Electronic Density of States*, Science **321**, 554 (2008).

[11] W. Liu, X. Tan, K. Yin, H. Liu, X. Tang, J. Shi, Q. Zhang, and C. Uher, *Convergence of Conduction bands as a Means of Enhancing Thermoelectric performance n-type of $Mg_2Si_{1-x}Sn_x$ Solid Solutions* Phys. Rev. Lett. **108**, 166601 (2012).

[12] S. Perumal, M. Samanta, T. Ghosh, U. S. Shenoy, A. K. Bohra, S. Bhattacharya, A. Singh, U. V. Waghmare, and K. Biswas, *Realization of High Thermoelectric Figure of Merit in GeTe by Complementary Co-Doping of Bi and In*, Joule **3**, 2565 (2019).

[13] P. Graziosi and N. Neophytou, *The Role of Electronic Bandstructure Shape in Improving the Thermoelectric Power Factor of Complex Materials*, ACS Appl. Electron. Mater. **6**, 2889 (2024).

[14] P. Graziosi and N. Neophytou, *Ultra-High Thermoelectric Power Factors in Narrow Gap Materials with Asymmetric Bands*, J. Phys. Chem. C **124**, 18462 (2020).

[15] E. Witkoske, X. Wang, M. Lundstrom, V. Askarpour, and J. Maassen, *Thermoelectric Band Engineering: The Role of Carrier Scattering*, J. Appl. Phys. **122**, 175102 (2017).

[16] C. Kumarasinghe and N. Neophytou, *Band Alignment and Scattering Considerations for Enhancing the Thermoelectric Power Factor of Complex Materials: The Case of Co-Based Half-Heusler Alloys*, Phys. Rev. B **99**, 195202 (2019).

[17] J. Park, M. Dylla, Y. Xia, M. Wood, G. J. Snyder, and A. Jain, *When Band Convergence Is Not Beneficial for Thermoelectrics*, Nat. Commun. **12**, 3425 (2021).

[18] V. Askarpour and J. Maassen, *First-Principles Analysis of Intravalley and Intervalley Electron-Phonon Scattering in Thermoelectric Materials*, Phys. Rev. B **107**, 045203 (2023).

[19] M. Thesberg, H. Kosina, and N. Neophytou, *On the Lorenz Number of Multiband Materials*, Phys. Rev. B **95**, 125206 (2017).

[20] J. J. Gong, A. J. Hong, J. Shuai, L. Li, Z. B. Yan, Z. F. Ren, and J. M. Liu, *Investigation of the Bipolar Effect in the Thermoelectric Material $CaMg_2Bi_2$ Using a First-Principles Study*, Phys. Chem. Chem. Phys. **18**, 16566 (2016).

[21] Y. Liu et al., *A Wearable Real-Time Power Supply with a $Mg_3Bi_2$-Based Thermoelectric Module*, Cell Rep. Phys. Sci. **2**, 100412 (2021).

[22] Z. Liu, W. Gao, H. Oshima, K. Nagase, C. H. Lee, and T. Mori, *Maximizing the Performance of N-Type $Mg_3Bi_2$ Based Materials for Room-Temperature Power Generation and Thermoelectric Cooling*, Nat. Commun. **13**, 1120 (2022).

[23] J. Yang et al., *Next-Generation Thermoelectric Cooling Modules Based on High-Performance $Mg_3(Bi,Sb)_2$ Material*, Joule **6**, 193 (2022).

[24] P. Graziosi, Z. Li, and N. Neophytou, *Bipolar Conduction Asymmetries Lead to Ultra-High Thermoelectric Power Factor*, Appl. Phys. Lett. **120**, 072102 (2022).

[25] C. Fu, Y. Sun, and C. Felser, *Topological thermoelectrics*, APL Mater. **8**, 040913 (2020).

[26] H. Karamitaheri, N. Neophytou, M. Pourfath, R. Faez, and H. Kosina, *Engineering Enhanced Thermoelectric Properties in Zigzag Graphene Nanoribbons*, J. Appl. Phys. **111**, 054501 (2012).

[27] A. Masci, E. Dimaggio, N. Neophytou, D. Narducci, G. Pennelli, *Large increase of the thermoelectric power factor in multi-barrier nanodevices*, Nano Energy, **132**, 110391, (2024).

[28] T. Ishibe, A. Tomeda, K. Watanabe, Y. Kamakura, N. Mori, N. Naruse, Y. Mera, Y. Yamashita, Y. Nakamura, *Methodology of Thermoelectric Power Factor Enhancement by Controlling Nanowire Interface*, ACS Appl. Mater. Interfaces **10**, 37709 (2018).

[29] Y. Liu, D. Cadavid, M. Ibáñez, S. Ortega, S. Martí-Sánchez, O. Dobrozhan, M. V. Kovalenko; J. Arbiol, and A. Cabot, *Thermoelectric properties of semiconductor-metal composites produced by particle blending,* APL Mater., **4**, 104813 (2016).

[30] M. Zebarjadi, K. Esfarjani, M. S. Dresselhaus, Z. F. Ren and G. Chen, *Perspectives on thermoelectrics: from fundamentals to device applications*, Energy Environ. Sci., **5**, 5147-5162, (2012).

[31] N. Neophytou, S. Foster, V. Vargiamidis, G. Pennelli, and D. Narducci, *Nanostructured Potential Well/Barrier Engineering for Realizing Unprecedentedly Large Thermoelectric Power Factors*, Materials Today Physics **11**, 100159 (2019).

[32] J. P. Heremans, B. Wiendlocha, and A. M. Chamoire, *Resonant Levels in Bulk Thermoelectric Semiconductors*, Energy Environ. Sci. **5**, 5510, (2012).

[33] F. Garmroudi, M. Parzer, A. Riss, C. Bourges, S. Khmelevskyi, T. Mori, E. Bauer, and A. Pustogow. *High Thermoelectric Performance in Metallic NiAu Alloys via Interband Scattering*, Sci. Adv. **9**, eadj1611 (2023).

[34] A. Riss, F. Garmroudi, M. Parzer, C. Eisenmenger-Sittner, A. Pustogow, T. Mori, and E. Bauer, *Material-efficient preparation and thermoelectric properties of metallic $Ni_xAu_{1-x}$ films with large power factor,* Phys. Rev. Materials 8, 095403, (2024).

[35] D. M. Rowe, G. Min and V. L. Kuznestsov, *Electrical resistivity and Seebeck coefficient of hot-pressed YbAl3 over the temperature range 150± 700 K,* Philosophical Magazine Letters, **77**, 2, 105, (1998).

[36] E. Bauer, 'Anomalous properties of Ce-Cu- and Yb-Cu-based compounds,' Advances in Physics, **40**, 4, 417-534, (1991).

[37] D. M. Rowe, V. L. Kuznetsov, L. A. Kuznetsova, and G. Min, *Electrical and thermal transport properties of intermediate-valence YbAl3,* J. Phys. D: Appl. Phys. **35**, 2183–2186, (2002).

[38] K.J. Proctor, C.D.W. Jones, F.J. DiSalvo, *Modification of the thermoelectric properties of CePd3 by the substitution of neodymium and thorium,* Journal of Physics and Chemistry of Solids **60**, 663–671, (1999).

[39] J. Yang, L. Xi, W. Qiu, L. Wu, X. Shi, L. Chen, J. Yang, W. Zhang, C. Uher, and D. J. Singh, *On the Tuning of Electrical and Thermal Transport in Thermoelectrics: An Integrated Theory-Experiment Perspective*, npj Computational Materials, **2**, 15015 (2016).

[40] P. Graziosi, C. Kumarasinghe, and N. Neophytou, *Impact of the Scattering Physics on the Power Factor of Complex Thermoelectric Materials*, J. Appl. Phys. **126**, 155701 (2019).

[41] J. Park, Y. Xia, V. Ozoliņš, and A. Jain, *Optimal Band Structure for Thermoelectrics with Realistic Scattering and Bands*, npj Comput. Mater. **7**, 43 (2021).

[42] J. Maassen, *Limits of Thermoelectric Performance with a Bounded Transport Distribution*, Phys. Rev. B **104**, 184301 (2021).

[43] J. M. Ziman, *Electrons and Phonons* (Cambridge, 1960).

[44] J. B. Goodenough, *Band Structure of Transition Metals and Their Alloys*, Phys. Rev. **120**, 67 (1960).

[45] E. B. Isaacs and C. Wolverton, *Remarkable Thermoelectric Performance in $BaPdS_2$ via Pudding-Mold Band Structure, Band Convergence, and Ultralow Lattice Thermal Conductivity*, Phys. Rev. Mater. **3**, 015403 (2019).

[46] Y. Pan et al., *Thermoelectric Properties of Novel Semimetals: A Case Study of $YbMnSb_2$*, Adv. Mater. **33**, 2003168 (2021).

[47] B. Hinterleitner et al., *Thermoelectric Performance of a Metastable Thin-Film Heusler Alloy*, Nature **576**, 85 (2019).

[48] T. Graf, C. Felser, and S. S. P. Parkin, *Simple Rules for the Understanding of Heusler Compounds*, Progress in Solid State Chemistry **39**, 1 (2011).

[49] F. Garmroudi, M. Parzer, A. Riss, S. Beyer, S. Khmelevskyi, T. Mori, M. Reticcioli, and E. Bauer, *Large Thermoelectric Power Factors by Opening the Band Gap in Semimetallic Heusler Alloys*, Materials Today Physics **27**, 100742 (2022).

[50] K. Durczewski and M. Ausloos, *Theory of the Thermoelectric Power of Model Semimetals and Semiconductors*, Z. Phys. B Condensed Matter **85**, 59 (1991).

[51] See Supplemental Material at [URL will be inserted by publisher] for indications of the Seebeck coefficient of many realistic materials, and information about the scattering rates in our simulations, the effect of intra- versus inter-valley scattering in degenerate VBs, band schematics for the narrow band cases in Fig. 6, optical phonon energies for various materials, and the effects of band tail broadening around the band edges.

[52] N. Neophytou, *Theory and Simulation Methods for Electronic and Phononic Transport in Thermoelectric Materials,* Springer Briefs in Physics, (2020).

[53] M. Lundstrom, *Fundamentals of Carrier Transport*, 2nd edn. Cambridge University Press, Cambridge (2000).

[54] Y. Jin, X. Wang, M. Yao, D. Qiu, D. J. Singh, J. Xi, J. Yang, and L. Xi, *High-throughput deformation potential and electrical transport calculations,* npj Computational Materials, **9**, 190 (2023).

[55] J. Zhou, H. Zhu, T. H. Liu, Q. Song, R. He, J. Mao, Z. Liu, W. Ren, B. Liao, D. J. Singh, Z. Ren, G. Chen, *Large thermoelectric power factor from crystal symmetry protected non-bonding orbitals in half Heuslers,* Nat. Commun. **9**, 1721 (2018).

[56] A. Franceschetti, S. H. Wei, and A. Zunger, *Absolute deformation potentials of Si, Al, NaCl,* Phys. Rev. B, **50**, 24 (1994).

[57] Z. Li, P. Graziosi, and N. Neophytou, *Efficient First-Principles Electronic Transport Approach to Complex Band Structure Materials: The Case of n-Type Mg$_3$Sb$_2$*, npj Comput. Mater. **10**, 8 (2024).

[58] J. Cao, J. D. Querales-Flores, A. R. Murphy, S. Fahy, and I. Savić, D*ominant electron-phonon scattering mechanisms in n-type PbTe from first principles,* Phys. Rev. B **98**, 205202, (2018).

[59] Z. Li, P. Graziosi, N. Neophytou, *Deformation potential extraction and computationally efficient mobility calculations in silicon from first principles,* Phys. Rev. B, **104,** 195201 (2021).

[60] C B So, K Takegahara, S. Wang, *Electronic structure of metals: V. Density of states, effective mass of the alkali metals,* J. Phys. F : Metal Physics, **7**, 8 (1977).

[61] O. K. Anderson, A. R. Mackintosh, *Fermi surfaces and effective masses in F.C.C. transition metals*, Solid State Communications, **6**, 285-290 (1960).

[62] N. W. Ashcroft, and N. D. Mermin, *Solid State Physics*, (1976).

[63] M. Mitra, A. Benton, Md. S. Akhanda, J. Qi, M. Zebarjadi, D. J. Singh, S. J. Poon, *Conventional Half-Heusler alloys advance* state-of-the-art *thermoelectric properties.* Materials Today Physics, **28**, 100900 (2022).

[64] D. I. Bilc, G. Hautier, D. Waroquiers, G. M. Rignanese, and P. Ghosez, *Low-Dimensional Transport and Large Thermoelectric Power Factors in Bulk Semiconductors by Band Engineering of Highly Directional Electronic States,* Phys. Rev. Lett. **114**, 136601 (2015).

[65] D. Bourgault, H. Hajoum, S. Pairis, O. Leynaud, R. Haettel, J. F. Motte, O. Rouleau, E. Alleno, *Improved Power factor in self-substituted Fe2VAl Thermoelectric thin films prepared by Co-sputtering,* ACS Appl, Energy Mater. **6**, 3 1526-1532 (2023).

[66] P. Graziosi, C. Kumarasinghe, and N. Neophytou, *Material Descriptors for the Discovery of Efficient Thermoelectrics*, ACS Appl Energy Mater **3**, 5913 (2020).

[67] R. D'Souza, J. Cao, J. D. Q. Flores, S. Fahy, I. Savic, *Electron-phonon scattering and thermoelectric transport of PbTe from first principles,* Phys, Rev. B., **102**, 115204, (2020).

[68] Y. Saberi, S. A. Sajjadi, *Comprehensive review on the effects of doing process on the thermoelectric properties of Bi2Te3 based alloys.* J. Alloys Compd. **904**, 163918, (2022).

[69] J.S. Young, and R. G. Reddy, *Processing and Thermoelectric Properties of TiNiSn Materials: A Review,* J. Mater. Eng. Performance, **28**, 5917-5930 (2019).

[70] M. Wolf, R. Hinterding, and A. Feldhoff, *High Power Factor vs High-ZT- A Review of Thermoelectric Materials For High Temperature Application,* Entropy, **21(11)** 1058 (2019).

[71] H. Zhu, W. Li, A. Nozariasbmarz, N. Liu, Y. Zhang, S. Priya, and B. Poudel, *Half-Heusler alloys as emerging high power density thermoelectric cooling materials,* Nat. Comm. , **14**, 3300 (2023).

[72] B. Xu, M. D. Gennaro, and M. J. Verstraete, *Thermoelectric properties of elemental metals from first principles electron-phonon coupling,* Phys. Rev. B. **102,** 155128 (2020).

[73] C. A. Domencali, and F. A. Otter, *Thermoelectric Power and Electron Scattering in Metal alloys,* Phys. Rev. **95**, 5 (1954).

[74] J. Mao, Y. Wang, H. S. Kim, Z. Liu, U. Saparamadu, F. Tian, K. Dahal, J. Sun, S. Chen, W. Liu, Z. Ren, *High thermoelectric power factor in Cu–Ni alloy originate from potential barrier scattering of twin boundaries. Nano Energy* **17**, 279–289 (2015).

[75] S. Li, K. Snyder, M. S. Akhanda, R. Martukanitz, M. Mitra, J. Poon, M. Zebarjadi, *Cost-efficient copper-nickel alloy for active cooling applications*. Int. J. Heat Mass Transf. **195**, 123181 (2022).

[76] C. Ho, R. Bogaard, T. Chi, T. Havill, H. James, *Thermoelectric power of selected metals and binary alloy systems*. Thermochimica Acta **218**, 29–56 (1993).

[77] V. L. Moruzzi and J. F. Janak, *Calculated thermal properties of metals,* Phys. Rev. B, **37**, 2, 790, (1988).

[78] A. Fernandez Guillermet and G. Grimvall, *Homology of interatomic forces and Debye temperatures in transition metals,* Phys. Rev. B, **40**, 3, 1521, (1989).

[79] Ziman, J. M. (1972). *Principles of the Theory of Solids*.

[80] G. Kaiblinger-Grujin, H. Kosina, and S. Selberherr, *Influence of the doping element on the electron mobility in n-silicon*, J. Appl. Phys. **83**, 3096–3101 (1998).

[81] J. Gull and Hans Kosina, *Monte Carlo study of electron–electron scattering effects in FET channels*, Solid-State Electronics **208**, 108730 (2023).

[82] V. Vargiamidis and N. Neophytou, *Hierarchical Nanostructuring Approaches for Thermoelectric Materials with High Power Factors*, Phys. Rev. B **99**, 045405 (2019).

[83] V. Vargiamidis, M. Thesberg, and N. Neophytou, *Theoretical Model for the Seebeck Coefficient in Superlattice Materials with Energy Relaxation*, J. Appl. Phys. **126**, 055105 (2019).

[84] H. Nautiyal and P. Scardi, *Thermoelectric Properties and Thermal Transport in Two-Dimensional $GaInSe_3$ and $GaInTe_3$ Monolayers: A First-Principles Study*, J. Appl. Phys. **135**, 174301 (2024).

[85] Y. Nishino, *Electronic Structure and Transport Properties of Pseudogap System $Fe_2VAl$*, Material Transactions **42**, 902 (2001).

[86] M. E. Jamer, B. A. Assaf, T. Devakul, and D. Heiman, *Magnetic and Transport Properties of $Mn_2CoAl$ Oriented Films*, Appl. Phys. Lett. **103**, 142403 (2013).

[87] M. E. Jamer, L. G. Marshall, G. E. Sterbinsky, L. H. Lewis, and D. Heiman, *Low-Moment Ferrimagnetic Phase of the Heusler Compound $Cr_2CoAl$*, J. Magn. Magn. Mater. **394**, 32 (2015).

[88] M. E. Jamer et al., *Compensated Ferrimagnetism in the Zero-Moment Heusler Alloy $Mn_3Al$*, Phys. Rev. Appl. **7**, 064036 (2017).

[89] S. R. Hari, V. Srinivas, C. R. Li, and Y. K. Kuo, *Thermoelectric Properties of Rare-Earth Doped $Fe_2VAl$ Heusler Alloys*, Journal of Physics: Condensed Matter **32**, 355706 (2020).

[90] R. Dutt, D. Pandey, and A. Chakrabarti, *Probing the Martensite Transition and Thermoelectric Properties of $Co_xTaZ$ (Z = Si, Ge, Sn and x = 1, 2): A Study Based on Density Functional Theory*, Journal of Physics: Condensed Matter **33**, 045402 (2021).

[91] A. Ślebarski, M. Fijałkowski, J. Deniszczyk, M. M. Maśka, and D. Kaczorowski, *Off-Stoichiometric Effect on Magnetic and Electron Transport Properties of $Fe_2VAl_{1.35}$ and $Ni_2VAl$: A Comparative Study*, Phys. Rev. B **109**, 165105 (2024).

[92] A. Nakano, A. Yamakage, U. Maruoka, H. Taniguchi, Y. Yasui, I. Terasaki, *Giant Peltier conductivity in an uncompensated semimetal $Ta_2PdSe_6$*. J. Phys. Energy, **3**, 044004 (2021).

[93] K. Kuga, K. Hirata, M. Matsunami, T. Takeuchi, *Huge Peltier conductivity in valence fluctuating material $Yb_3Si_5$*. Appl. Phys. Lett. **123**, 202201 (2023).

[94] M. E. Jamer, B. A. Assaf, G. E. Sterbinsky, D. Arena, L. H. Lewis, A. A. Saúl, G. Radtke, and D. Heiman, *Antiferromagnetic phase of the gapless semiconductor $V_3Al$*, Phys. Rev. B **91**, 094409 (2015).

[95] V. Kanchan, G. Vaitheeswaran, Y. Ma, Y. Xie, A. Svane, and O. Eriksson, *Density functional study of elastic and vibrational properties of the Heusler alloys, Fe2VAl, Fe2VGa,* Phys, Rev. B. **80**, 125108 (2009).

[96] S. A. Barczak, R. J. Quinn, J. E. Halpin, K. Domosud, R. I. Smith, A. R. Baker, E. Don, I. Forbes, K. Refson, D. A. MacLaren, J. W. G. Bos, *Suppression of thermal conductivity without impeding electron mobility in n-type XNiSn half-Heusler thermoelectrics.* J. Mat. Chem. A **7**, 27124 (2019).

[97] N. F. Mott and E. A. Davis, Electronic Processes in Non-crystalline Materials, Clarendon, Oxford, 1971), p. 47.

[98] L. D. Hicks and M. S. Dresselhaus, *Thermoelectric figure of merit of a one-dimensional conductor,* Phys. Rev. B 47, 16631(R), 1993.

[99] P. Priyadarshi, V. Vargiamidis and N. Neophytou, *Energy Filtering in Doping Modulated Nanoengineered Thermoelectric Materials: A Monte Carlo Simulation Approach,* Materials **17**(14), 3522, (2024).

[100] G. D. Mahan, J.O. Sofo, *The best thermoelectric*, Proc. Natl. Acad. Sci. USA, .**93**, 7436 (1996).

[101] C. Jeong, R. Kim, M. S. Lundstrom, *On the best bandstructure for thermoelectric performance: A Landauer perspective*, J. Appl. Phys. **111**, 113707 (2012).

[102] Z. Tong, S. Li, X. Ruan, and Hua Bao, *Comprehensive first principles analysis of phonon thermal conductivity, and electron-phonon coupling in different metals,* Phys. Rev. B, **100**, 144306 (2019).

[103] R. L. Powell, W. J. Hall, H. M. Roder, *Low-Temperature Transport Properties of Commercial Metals and Alloys. II. Aluminums*, J. Appl. Phys. **31**, 496–503 (1960).

# Materials design criteria for ultra-high thermoelectric power factors in metals

Patrizio Graziosi, Kim-Isabelle Mehnert, Rajeev Dutt, Jan-Willem G. Bos, and Neophytos Neophytou

## Supplementary Information

**Section S1: Seebeck coefficient indicator**

Possible material systems to exploit the energy filtering of carriers promoted by inter-band scattering from the Materials Project database are provided, chosen among the ones identified as stable (not decomposing). The bandstructure and DOS are taken as images from the Materials Project website (https://next-gen.materialsproject.org/).

In order to provide an indication of the potential Seebeck coefficient improvements due to the sharp features that appear in the DOS, and promote enhanced scattering, we use BTE calculations and proceed as follows: we use Eq. (1) and (2) from the main text for the Seebeck coefficient:

$$\sigma_{ij(E_F,T)} = q_0^2 \int_E \Xi_{ij}(E)\left(-\frac{\partial f_0}{\partial E}\right) dE, \tag{S1}$$

$$S_{ij(E_F,T)} = \frac{q_0 k_B}{\sigma_{ij}} \int_E \Xi_{ij}(E)\left(-\frac{\partial f_0}{\partial E}\right)\frac{E-E_F}{k_B T} dE. \tag{S2}$$

For the TDF (Eq. 4 in the main text):

$$\Xi_{(E,T)} = \frac{1}{3} v_{(E)}^2 \tau_{(E)} g_{(E)} \tag{S3}$$

we need expressions for the velocity, the DOS, and the scattering times. We do not have access to those quantities, except the total DOS that we get from DFT. When we obtain the DOS from DFT for these metallic systems, the bands are heavily mixed, as shown in the multiple figures we provide below, and cannot be resolved individually, at least in a simplified manner. Thus, we cannot resolve their effective mass (*m\**), and cannot employ expressions that involve the effective mass directly as in the case of a parabolic band. The only option is to relate all three quantities to that total DOS (which involves many bands) in a sensible way, that would allow for the effect of inter-band scattering at the high DOS band regions. For this, we first assume that the scattering rates are proportional to the density of states, following from Fermi's Golden Rule, as standard practice – thus the scattering times are inversely proportional to the DOS, $g_{(E)}$. This captures the effect of scattering of the *s*-band carriers by the higher DOS *d*-bands, for example, as required. In the expression of the TDF then, $\tau_{(E)}$ and $g_{(E)}$ cancel out, and we are left with the TDF being proportional to $\Xi_{(E)} \sim v_{(E)}^2$. This approximation is exact for elastic isotropic scattering (ADP) and is still good enough at first order for optical phonon scattering (ODP) as well, especially for small values of the optical phonon energy involved in the scattering process.

We now need to relate $v_{(E)}^2$ to the total DOS as well. However, this also has to reflect the fact that when the DOS increases, scattering also increases and the conductivity decreases, i.e. in the energy region where the heavy bands overlap with the light bands, the conductivity of the light band suffers. Thus, any form of substitution we need to make, will need to have the DOS in the denominator to capture this. The bandstructures we deal with contain a mixture of bands, with different band edges in energy (thus we cannot relate directly the overall $v(E-E_0)^2 \sim g(E-E_0)^2$ since $E_0$ is different for different bands) and with different curvature, and inter-band scattering. Thus, one cannot manipulate the velocity and DOS of these systems to have an estimate between the velocity and DOS as we can do for the simple parabolic band relations that involve the effective masses.

Still, however, to replace $v_{(E)}^2$ we follow a simple approach by utilizing the fact that the velocity and DOS are in general adversely related with the band effective mass. Thus, we assume that an energy region with large DOS, will have bands with high effective mass, and thus low velocity. We then we still use the fact that in a parabolic band the velocity is $v \sim m^{-1/2}$ and the fact that the density of states DOS is $g \sim m^{3/2}$. Thus, after a simple manipulation, it turns out that $v \sim g^{-1/3}$. (We use this approximation, but note that at this simplified level one can use a TDF that is inversely proportional to the DOS at some exponent, i.e. if the scattering time is assumed constant and the velocity as inversely proportional to the DOS, etc.). Using this in the TDF, we use the BTE to extract an 'indicator' for the Seebeck coefficient. Note that here we assume parabolic approximations and remove the energy dependencies since it is difficult to resolve the band edge, in order to be able to provide a possible indication for the Seebeck behavior, and we avoid the tremendous difficulties required in exploring accurate transport in these materials. Thus, we only provide an 'indication' for the Seebeck, rather than actual values. What we use for the TDF in Eq. S2, $\Xi_{(E)} = Ag^{-2/3}$ (with $A$ being a constant that eventually drops out and with $v \sim g^{-1/3}$), is a simple approximation, but it is an approximation that makes sense, since it creates a TDF which is low in the high DOS, high scattering energy regions, and high in the low DOS, low scattering energy regions. To provide somewhat more relevance, the relation $v \sim g^{-1/3}$ essentially says that the (average) velocity is inversely proportional to the third root of the DOS. This is reflected to velocity being proportional to $v \sim dE/dk$, while the DOS being proportional in k-space to $g \sim dk^3$. Thus $v \sim g^{-1/3}$ makes the transport dictates by the velocity of the band states. Large DOS means that the effective mass is large, the velocity small and the conductivity small, and vice versa.

Alternatively, we can employ the Mott formula for the Seebeck coefficient directly [6], in which the Seebeck coefficient is proportional to the logarithmic derivative of the conductivity. In this case some approximations should also be considered since the only quantity we have access to is the total DOS from DFT. We assume that conductivity is degraded when the DOS is large, an indication of the larger scattering of the light band carriers into the heavier bands. Thus, we can consider that the conductivity is inversely proportional to the total DFT DOS and still use the Mott formula for a slightly different indication of the Seebeck coefficient. In general, we can actually consider that the TDFis proportional to the DOS raised to some negative exponent, as $\Xi_{(E)} = Ag^\alpha$, and in our case we have chosen $\alpha = -2/3$.

Using this, below we show several examples of metals with sharp increases in their DOS in some energy ranges, and plot the 'Seebeck indicator', which we denote with $S^*$. All DOS data are taken from the Materials Project database, and the material identifier number is also denoted. Clearly, when the DOS sharp features appear, peaks in the 'Seebeck indicator' also appear to values even up to $S^* \sim 50$ μV/K, similar to what we show in the main text, which could be significant in providing ultra-high PFs in these materials, provided that the Fermi level is placed in that vicinity. Note that in the case of magnetic materials the Seebeck coefficient is typically evaluated using the two channels method, with individual values corresponding to each spin and total Seebeck coefficient are combined as a weighted average of both spins [1, 2]. However, here our main focus is to probe the effect of change in DOS slope/values on the Seebeck coefficient. Hence, the results we present are based on the total density of states in those cases as well.

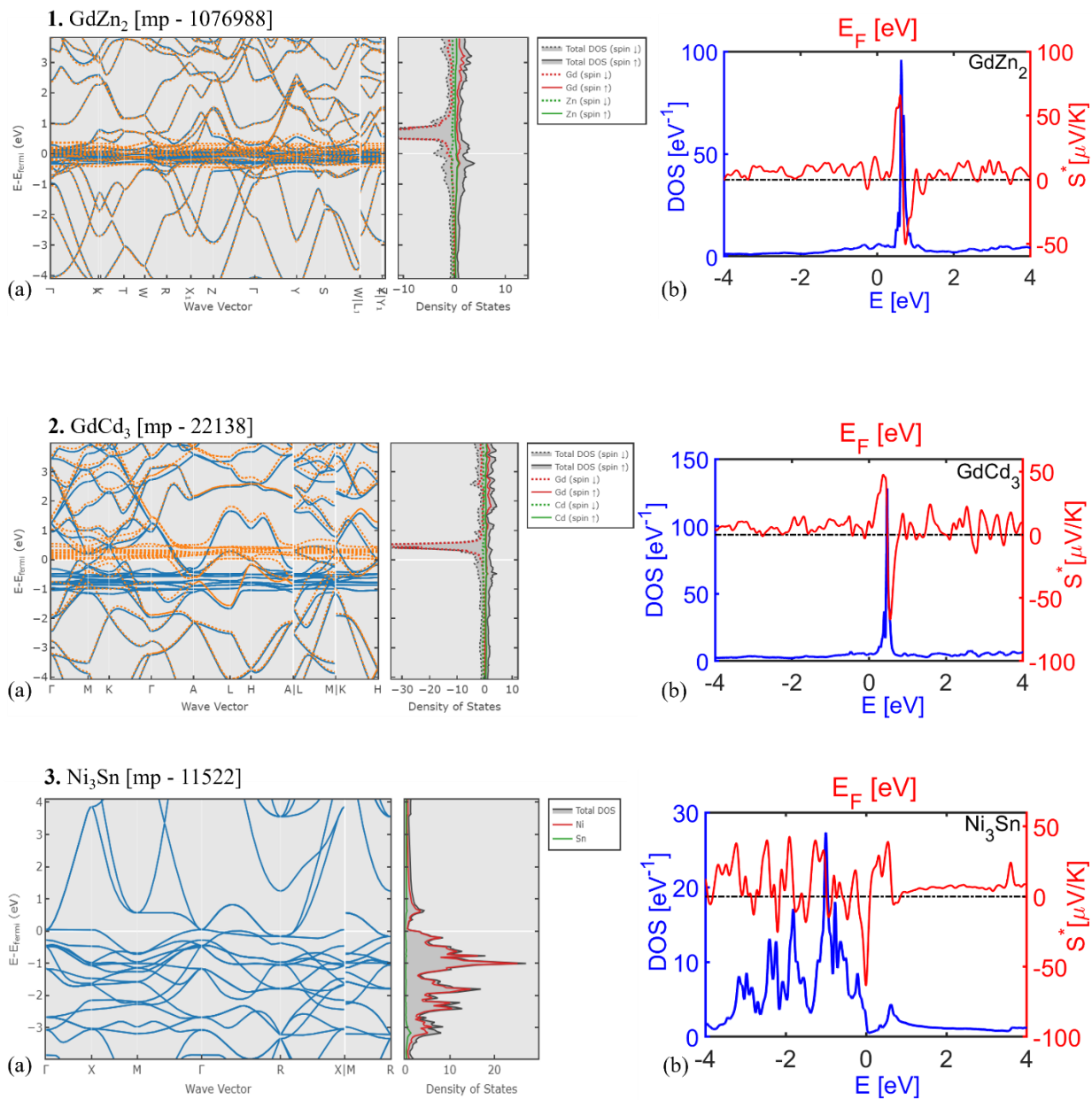

**4.** AlNi [mp - 1487]

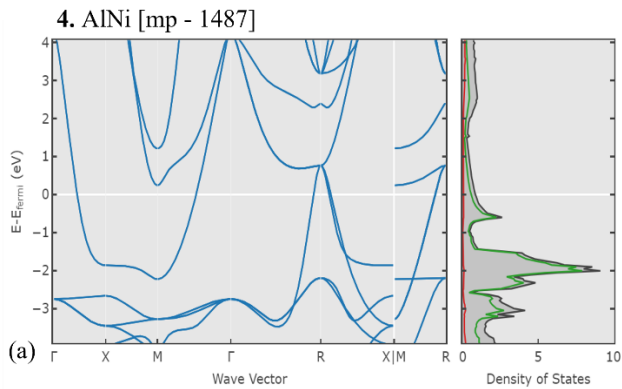
(a)

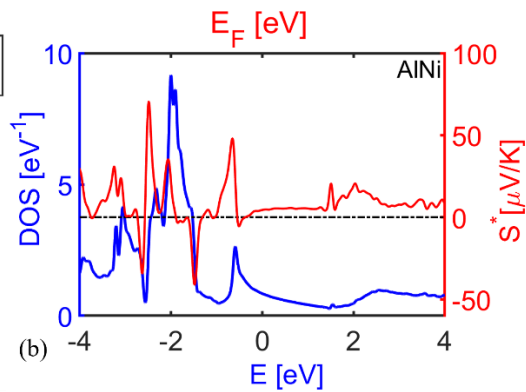
(b)

**5.** NiAu$_3$ [mp- 976784]

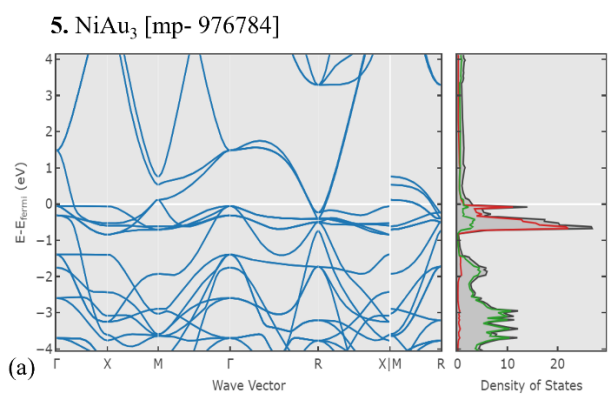
(a)

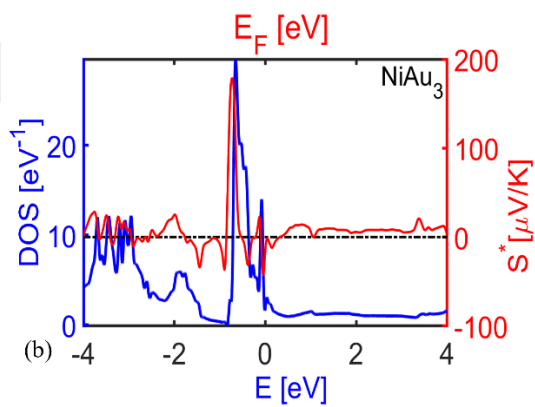
(b)

**6.** NiCu$_3$ [mp - 1225698]

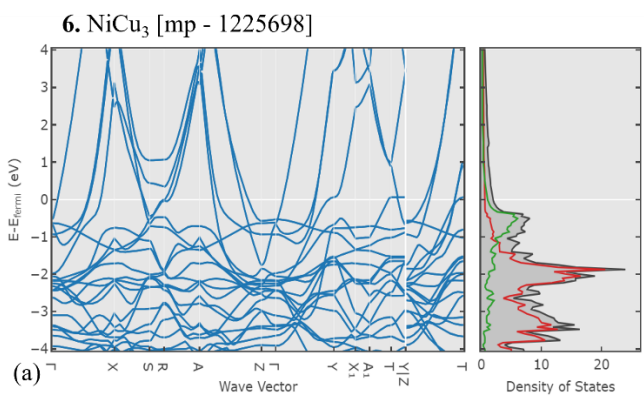
(a)

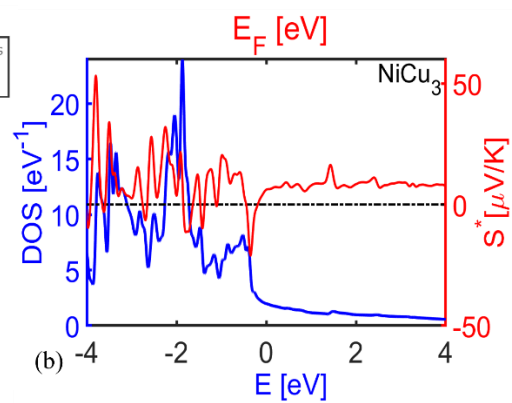
(b)

**7.** CoAu$_3$ [mp - 1206590]

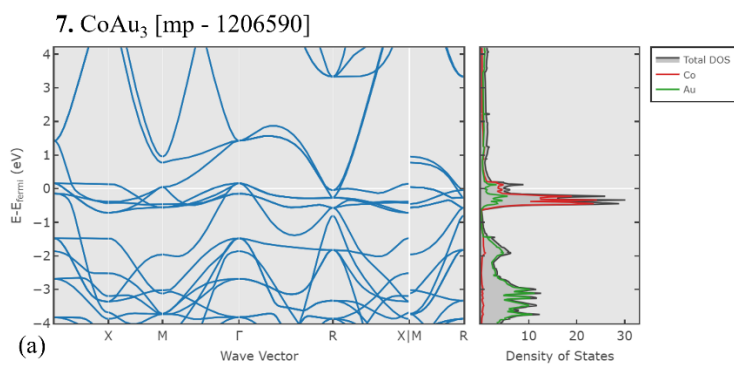
(a)

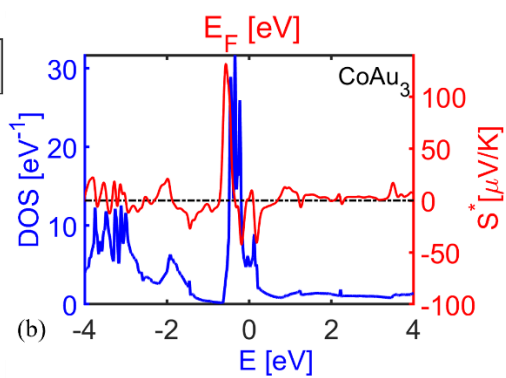
(b)

**8.** Ni$_3$Sb [mp - 976847]

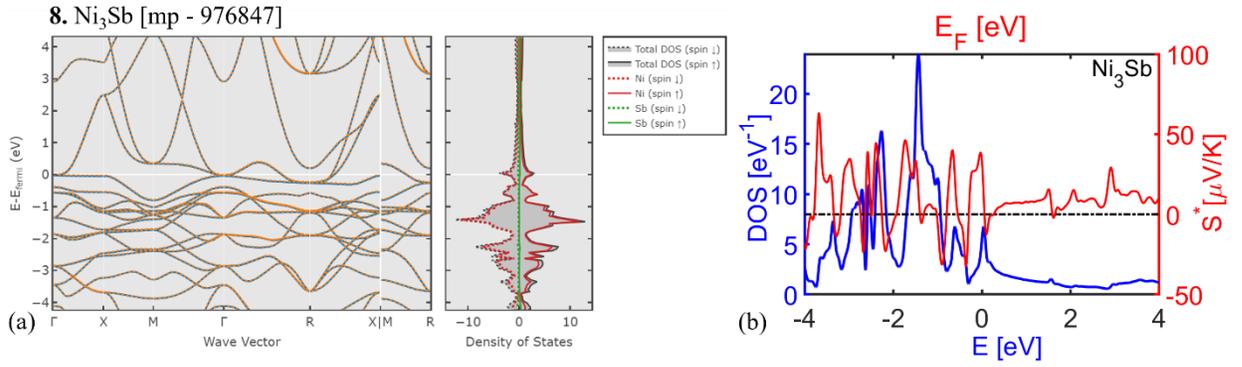

**9.** NiSb [mp - 810]

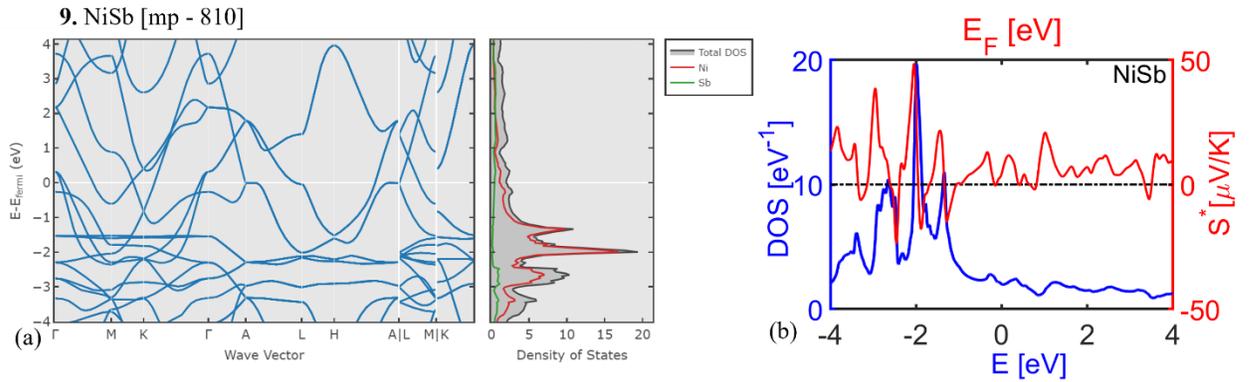

**10.** Ni$_3$Al [mp - 2593]

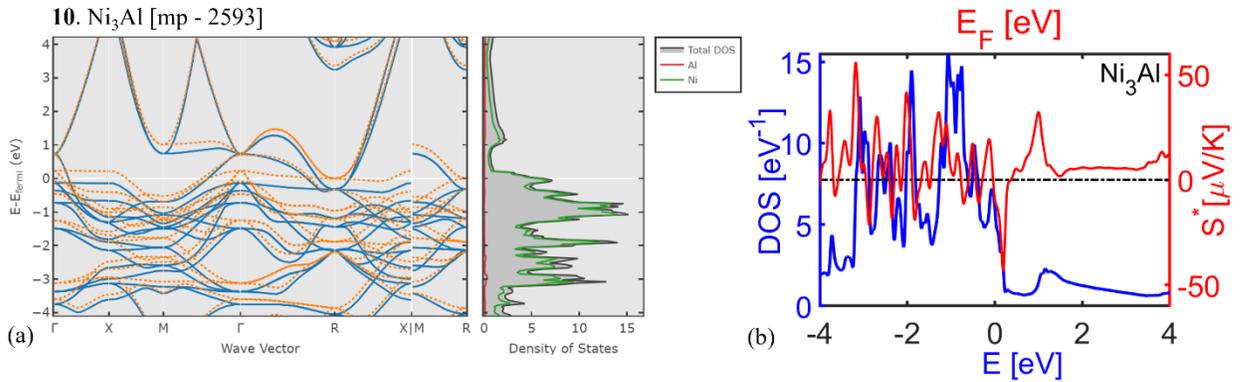

**11.** NiAl$_3$ [mp - 622209]

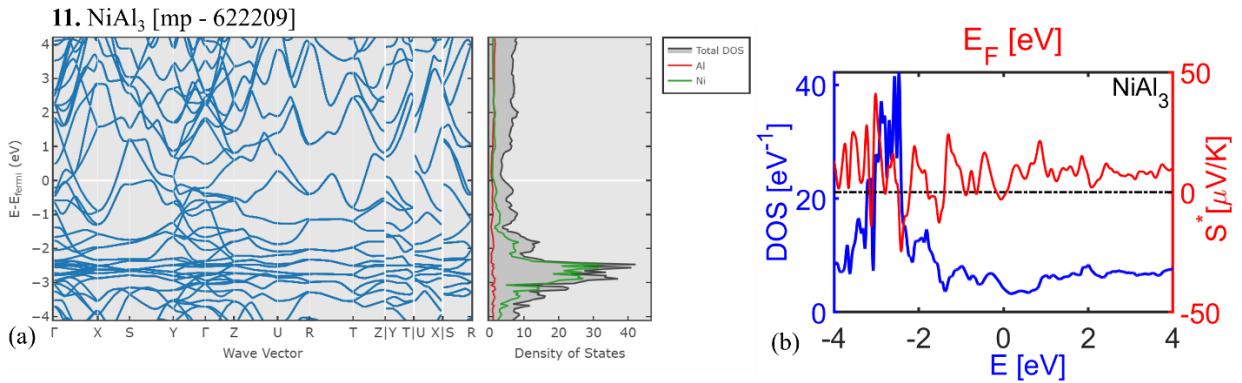

**12.** InNi [mp -20997]

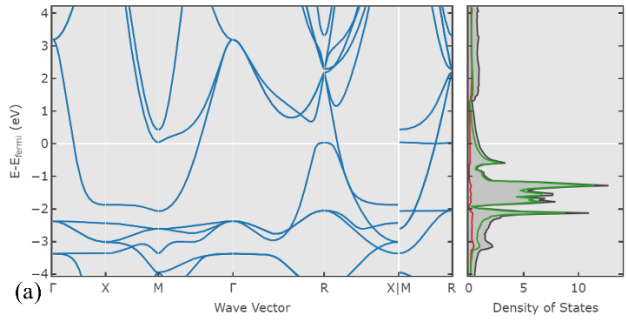
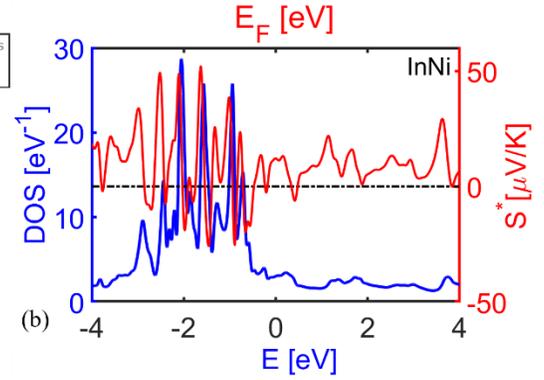

**13.** NiIn$_3$ [mp - 1206590]

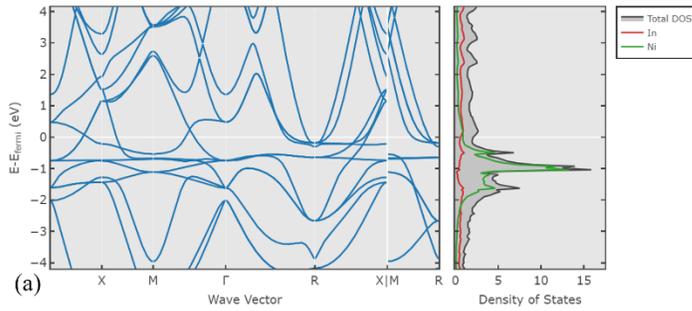
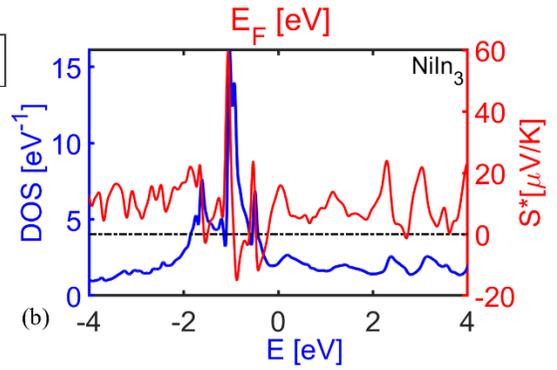

**14.** YbAl$_2$Ni [mp - 12782]

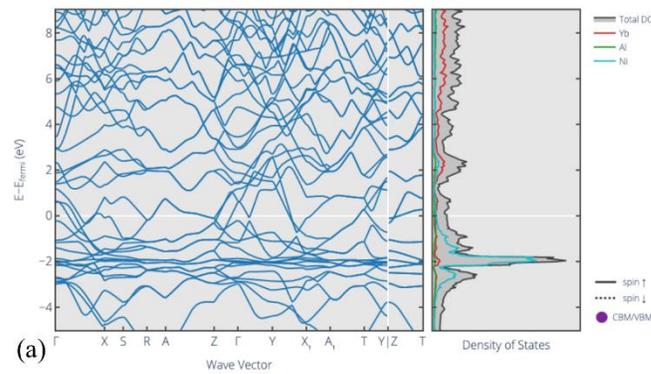
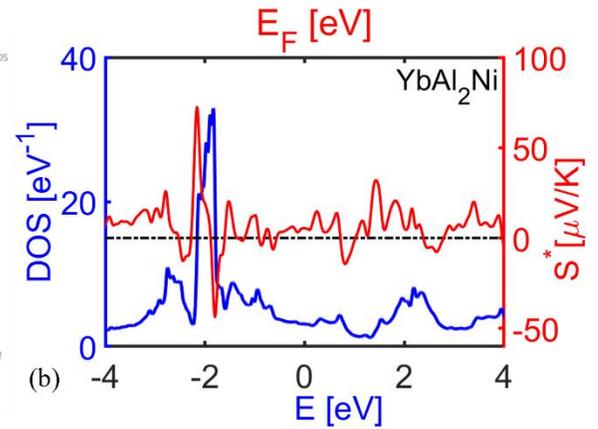

**15.** NiBi [mp - 1206590]

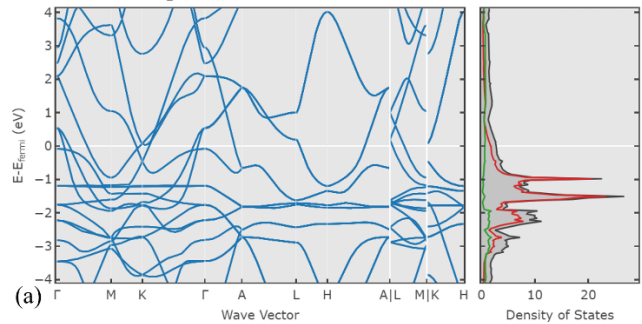
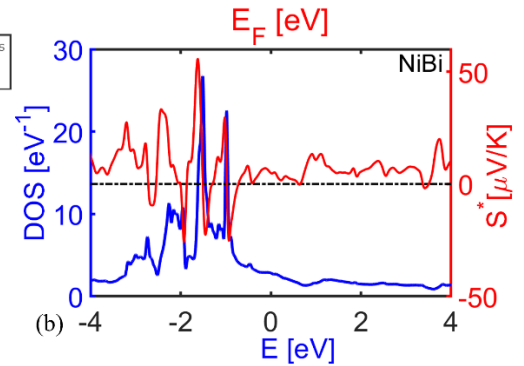

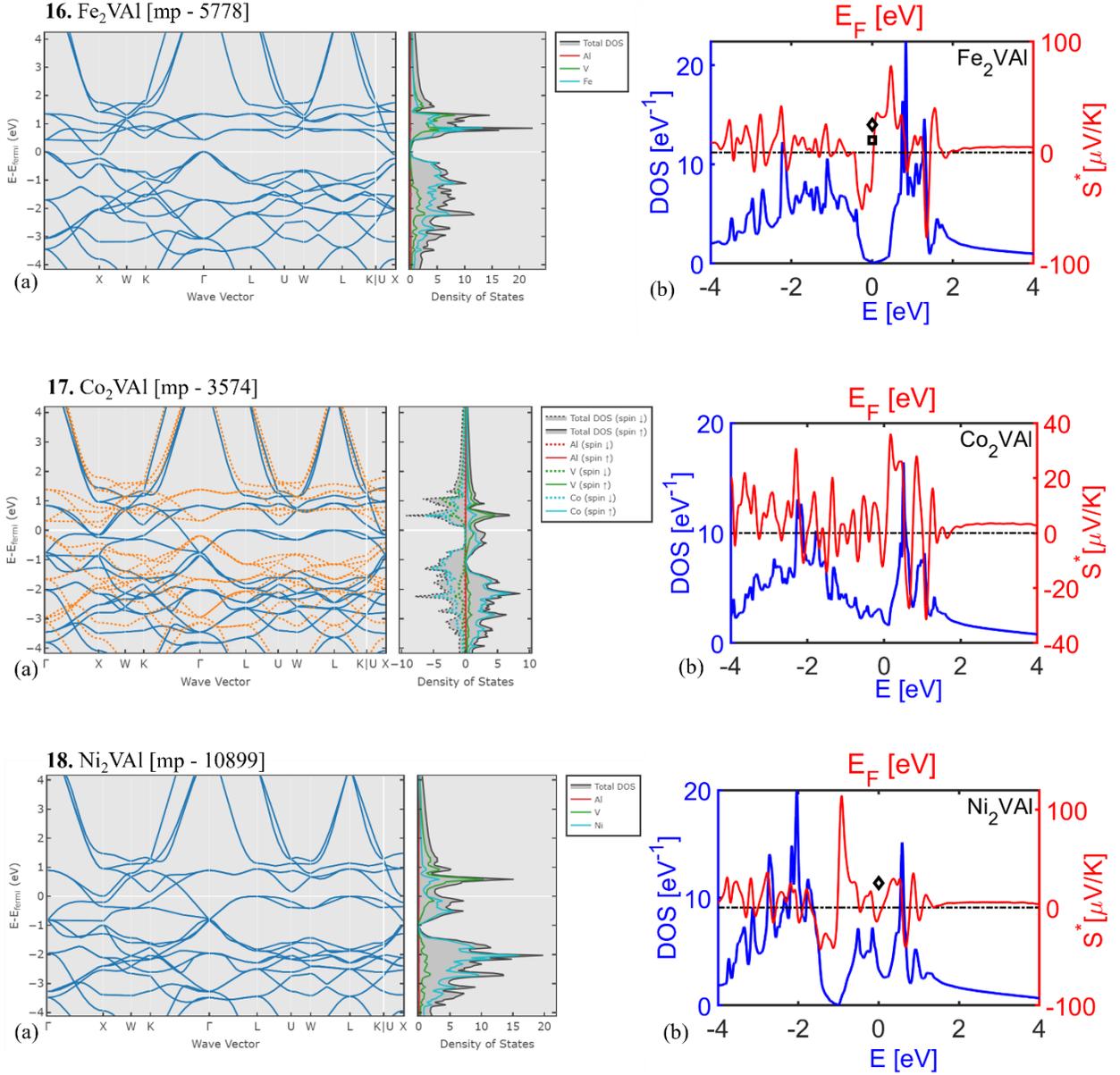

**Figure: S1.1-18**: Left panels (a): The electronic structure of the materials from the Material Project database and their DOS with their IDs denoted for reference. Right panels (b): The blue color line represents the total density of states of the corresponding system (same as in the left panels), and the red color line shows the calculated values of the 'Seebeck coefficient indicator' ($S^*$). The black symbols in some figures show the experimental values of the Seebeck coefficient (the diamond symbol in both $Fe_2VAl$ and $Ni_2VAl$ has been taken from Ref. [3], and the square symbol data point in $Fe_2VAl$ has been taken from Ref. [4]). These are the experimental values, but the exact value of carrier concentration is unknown, thus we have plotted these values at the Fermi level ($E = 0$ eV). In the case of $Fe_2VAl$, experimental evidence shows $S = 40$ μV/K in it's pure form. However, electronic rich and rare earth doping increase this value to the range of 100 μV/K, with change in sign and corroborate with above results [5].

References:

1. K. Vandaele, S. Watzman, B. Flebus, A. Prakash, Y. Zheng, S. R. Boona, J.P. Heremans, 1, 39-49 (2017).
2. R. Dutt, D. Pandey, A. Chakrabarti, J. Phys.: Condens. Matter 33, 045402 (2020)
3. A. Ślebarski, M. Fijałkowski, J. Deniszczyk, M. M. Maśka, and D. Kaczorowski Phys. Rev. B, 109, 165105 (2024).
4. Y. Nishino. Electronic, Magnetic and Transport Properties of the Pseudogap $Fe_2VAl$ system. *The Science of Complex Alloy Phases*, 325-344(2005).
5. S. R. Hari, V. Srinivas, C. R. Li and Y. K. Kuo, J. Phys.: Condens. Matter, 32, 355706 (2020)
6. N. F. Mott and E. A. Davis, Electronic Processes in Non-crystalline Materials, Clarendon, Oxford, 1971), p. 47.

## Section S2: scattering rates

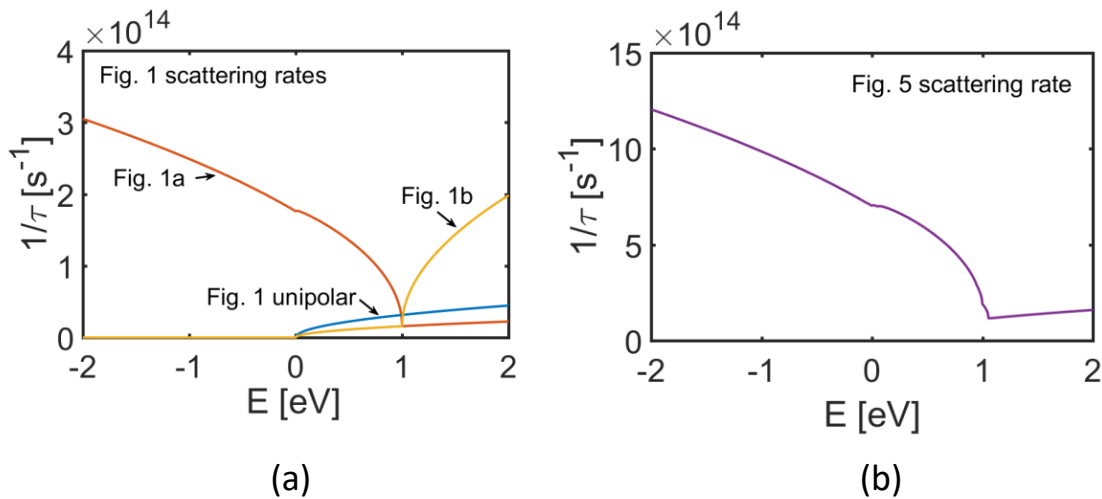

(a)        (b)

Figure S2.1: Scattering rates for the cases depicted in Fig. 1a-b and Fig. 5 (the stronger IVS) of the manuscript.

## Section S3: Intra- versus inter-valley scattering in the degenerate VBs

In the case of multiple light bands in the VB, if inter-valley scattering exists in addition to the intra-valley scattering in the VB, then it can be marginally beneficial to the PF, since it enhances the steepness of the transport distribution function and the steepness of the conductivity around the onset of the sharp feature. We show this in the figure below. **Figure S3.1** shows the TE coefficients for a VB with maxima at Γ, X, and W and a single valley CB. We include intra-valley ADP scattering for all valleys and ODP scattering for the inter-valley processes in the bands of the VB and between the CB-VB bands (purple lines). We then increase the ODP scattering strength for the inter-valley processes in the VB by doubling their deformation potential (blue line). A very narrow increase in the PF can be observed, though not significant. We then increase in addition the inter-band process between the CB and VB by doubling that deformation potential, and obtain the green line, which provides the larger PF improvement as this enhances the 'filtering-by-scattering' process.

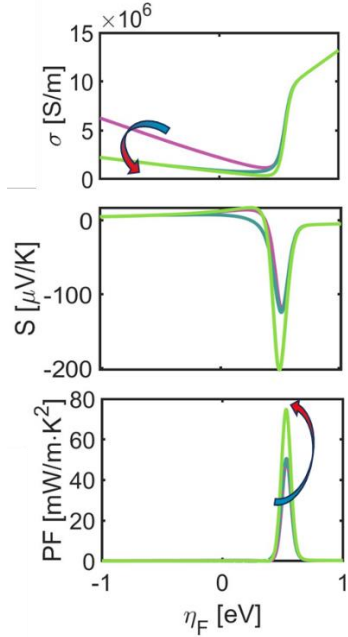

Figure S3.1: The TE coefficients conductivity, Seebeck coefficient and PF, respectively, for the bandstructure case consisting of a CB and multiple VBs. ADP intra-valley scattering and ODP inter-valley (between the VBs) and inter-band (between the CB and VBs) scattering is considered (purple line). Then the inter-valley scattering strength in the VB increases (blue line). Then the inter-band scattering strength from the VB to the CB increases in addition (green line).

## Section S4: Bands schematics for the narrow band cases of Fig. 6

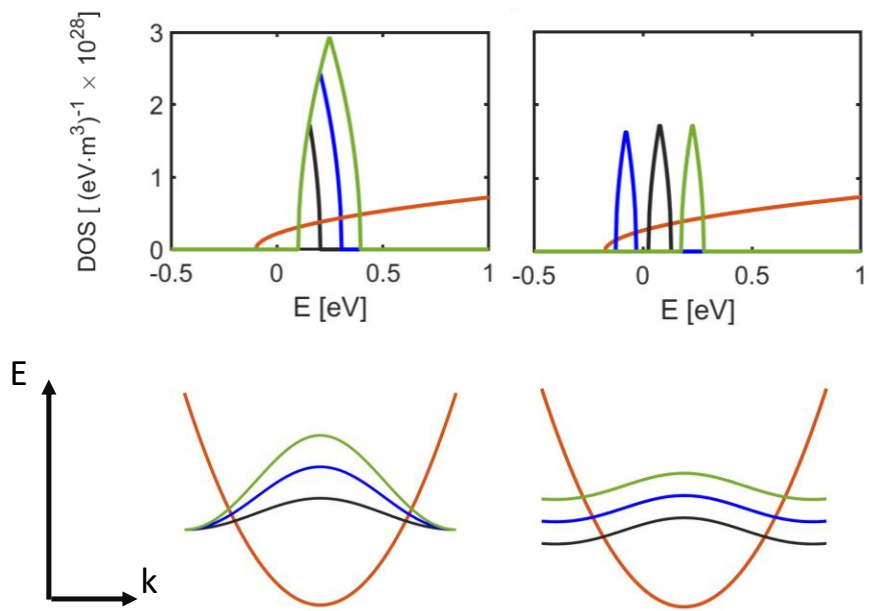

Figure S4.1: Schematics of the bandstructures for the narrow band cases in Fig. 6 of the main paper that correspond to the DOS of the two cases presented in the main paper. (The DOS figures are the same as in the main paper for reference.)

# Section S5: Optical phonon energies for materials (mainly metals)

| Material System | Phonon Energy of highest optical mode (meV) |
|---|---|
| $MgB_2$ | 100<br>10.1103/PhysRevLett.87.017005,<br>DOI:10.1038/s41598-017-03877-5 |
| Fe and Ni | 37<br>https://doi.org/10.1103/PhysRevB.62.273 |
| $Ni_3Sn$ | 31<br><br>[L. H. Li, W. L. Wang, and B. Wei, *First principle and molecular dynamics calculations for physical properties of Ni-Sn alloy system,* Comp. Mat. Sc., **99**, 274-284 (2015)]<br><br>https://doi.org/10.1016/j.commatsci.2014.11.031 |
| NiAl | 33<br>[Y. Wang, M. Liao, B. J. Bocklund, P. Gao, S. Li Shang, H. Kim, A. M. Beese, L. Q. Chen, Z. K., *DFTK: Density Functional Theory Toolkit for high-throughput lattice dynamics calculations,* Calphad, 75, 102355, (2021)]<br><br>https://www.sciencedirect.com/science/article/pii/S0364591621001024 |
| Al | 40<br>[S. Chantasiriwan, and F. Milstein, *Higher order elasticity of cubic metals in the Embedded-atom method*, Phys. Rev. B, **53** 21 (1996)] |
| Si | 60<br>[ S. Wei, M. Y. Chou, *Phonon dispersion of Si and Ge from first principles calculations,* Phys. Rev. B, **50**, 4 (1994)] |
| Ge | 38<br>[G. Neil, and G. Nelson, *phonon anharmonicity of Ge in range of 80-880K,* Phys. Rev. B, **10** 2 (1974)] |
| Be | 82<br>[R. Stedman, Z. Amilius, R. Pauli, and O Sundin *Phonon spectrum of Be at 80 K*, J. Phys. F: Met. Phys. **8,** 157 (1976)] |
| $NiCu_3$ | 29<br>[B. Onat, S. Durukanoglu *An optimized interatomic potential for Cu-Ni Alloys, with embedded-atom method*, J. Phys.: Cond. Matter, **26**, 035404 (2013)]<br>https://iopscience.iop.org/article/10.1088/0953-8984/26/3/035404 |
| $Ni_3Sb$ | 31 |

| | [O. G. Randl, W. Petry, G. Vogl, W. Buhrer, B Hennion, *Lattice dynamics and related diffusion properties of intermetalics: II Ni$_3$Sb,* J. Phys.: Cond. Matter, **9** 10283-10292 (1997)] https://iopscience.iop.org/article/10.1088/0953-8984/9/46/025/pdf |
|---|---|
| **Ni$_3$Al** | 37 [Y. Wang, M. Liao, B. J. Bocklund, P. Gao, S. Li Shang, H. Kim, A. M. Beese, L. Q. Chen, Z. K., *DFTK: Density Functional Theory Toolkit for high-throughput lattice dynamics calculations,* Calphad, 75, 102355, (2021)] https://www.sciencedirect.com/science/article/pii/S0364591621001024 |
| **NiAl$_3$** | 41 [Y. Wang, M. Liao, B. J. Bocklund, P. Gao, S. Li Shang, H. Kim, A. M. Beese, L. Q. Chen, Z. K., *DFTK: Density Functional Theory Toolkit for high-throughput lattice dynamics calculations,* Calphad, 75, 102355, (2021)] |
| **Fe$_2$VAl** | 45 [V. Kanchan, G. Vaitheeswaran, Y. Ma, Y. Xie, A. Svane, and O. Eriksson, *Density functional study of elastic and vibrational properties of the Heusler alloys, Fe2VAl, Fe2VGa,* Phys, Rev. B. **80**, 125108 (2009)] https://journals.aps.org/prb/pdf/10.1103/PhysRevB.80.125108 |

## Section S6: The effect of band tail broadening around the band edge

In reality non-idealities and disorder will introduce band tails and smearing of the sharp edges that are created in the vicinity of the heavy band edge, and are responsible for the large Seebeck coefficient increases. These non-idealities will potentially reduce the sharpness of the TDF around that region, and in that case, the large increases in the Seebeck coefficient and power factor (PF) we observe, will be mitigated. Still, since these effects have been observed experimentally as well, we don't expect the effect of band tails to be severe.

The purpose of this paper is to demonstrate an operating principle that can provide very large power factors, however, here we illustrate the effect of non-idealities, especially in the vicinity of the heavy band edge. For this, we use the based data from Fig. 2 of the main paper, and introduce narrow bands near the band edges, smeared our by Gaussian functions to resemble exponentially decaying states away from the and edge. In the example shown in Fig. S6.1 below, we have added bands that spread ~100 meV, with ~50 meV approximately in each side of the band edge, as shown by the DOS of Fig. S6.1a. We compare the transport coefficients in the original and disordered cases. The TDF becomes slightly broadened (blue line versus orange line in Fig. S6.1b), which reflects on the conductivity, Seebeck coefficient, and PF, which are reduced around the band edge. The reduction in the PF is of the order of 10% in this example. We find that the height of the disorder band does not have an effect on the PF, since it is typically overshadowed by the height of the heavy band. On the other hand, the more the spread of the disorder band is, the larger the reduction in the base PF. If we triple the disorder band spread to over 150meV above the heavy band edge, the PF drops to half the base PF value (blue line), which however, is still more than double compared to the PF of the single band, unipolar material.

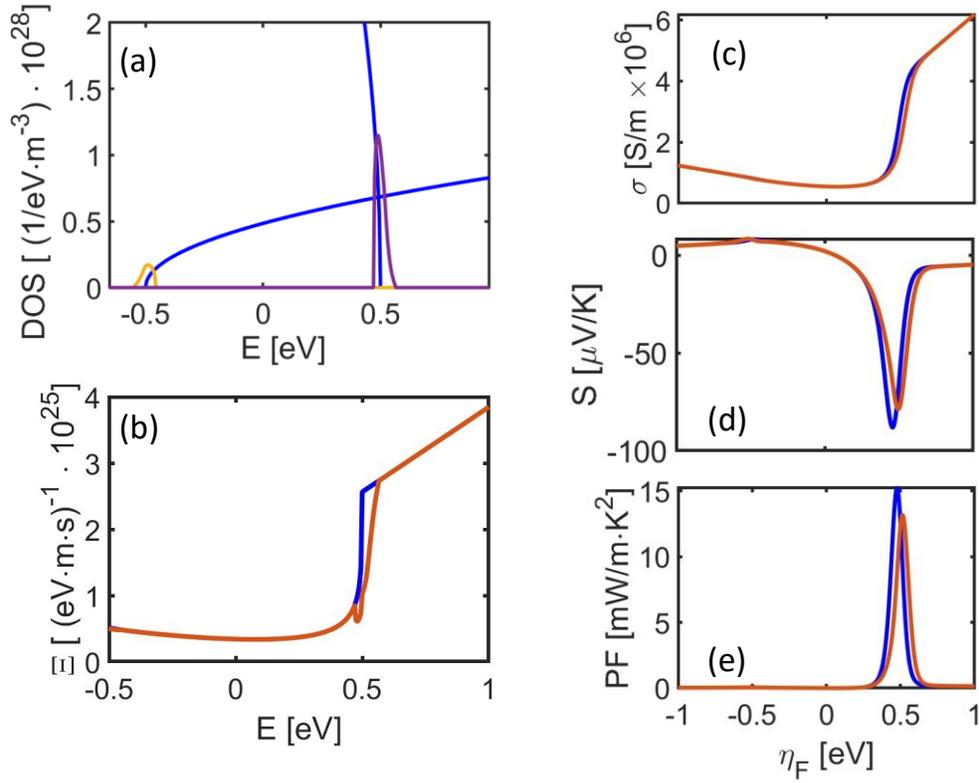

Figure S6.1: The example for the band overlap from Fig. 2 in the main paper, with Gaussian smearing disorder bands superimposed around the main band edges, representing impurity caused band tails and disorder. These extent for ~100 meV in energy width, with ~50 meV extension in each side of the band edge. (a) The DOS. (b) The TDF. (c-f) The TE coefficients electrical conductivity, Seebeck and PF.